\documentclass[pra,reprint,aps,superscriptaddress]{revtex4-2}
\usepackage[utf8]{inputenc}
\usepackage[T1]{fontenc}
\usepackage{epsfig}
\usepackage{mathptmx}
\usepackage{dcolumn}
\usepackage{bm}
\usepackage{siunitx}
\usepackage{graphicx}
\usepackage{times}
\usepackage{amssymb}
\usepackage{amsmath}
\usepackage{amsfonts}
\usepackage{enumitem} 
\usepackage{afterpage}
\usepackage{babel}
\usepackage{setspace}
\usepackage[dvipsnames]{xcolor}
\usepackage{titlesec}
\usepackage{upgreek}
\usepackage{mathtools}

\usepackage[sort&compress]{natbib}

\setcitestyle{open={},close={},numbers,comma,super}

\usepackage[pagebackref=false]{hyperref}
\hypersetup{
    colorlinks  =   true,
    linkcolor   =   BlueViolet,
    citecolor   =   BlueViolet,
    filecolor   =   BlueViolet,
    urlcolor    =   BlueViolet}
\usepackage[capitalise,nameinlink]{cleveref}
\usepackage{hypcap}

\usepackage{titlesec}
\titleformat{\section}[hang]{\small\bfseries\sffamily}{\thesection.}{0.5em}{}
\titlespacing{\section}{0pc}{1.2pc}{0.3pc}
\titlespacing{\subsection}{0pc}{1pc}{0.2pc}

\makeatletter
\renewcommand*{\fnum@figure}{{\normalfont\bfseries \figurename~\thefigure}}
\renewcommand*{\@caption@fignum@sep}{\textbf{ | }}
\renewcommand{\thetable}{\arabic{table}}
\renewcommand*{\fnum@table}{{\normalfont\bfseries \tablename~\thetable}}
\makeatother

\begin{document}

\title{A kilometer photonic link connecting superconducting circuits in two dilution refrigerators}

\author{Yiyu Zhou}
\author{Yufeng Wu}
\author{Chunzhen Li}
\author{Mohan Shen}
\author{Likai Yang}
\author{Jiacheng Xie}
\affiliation{Department of Electrical and Computer Engineering, Yale University, New Haven, Connecticut 06511, USA}

\author{Hong X. Tang}
\email[Corresponding author: ]{hong.tang@yale.edu}
\affiliation{Department of Electrical and Computer Engineering, Yale University, New Haven, Connecticut 06511, USA}
\affiliation{Department of Physics and Applied Physics, Yale University, New Haven, Connecticut 06511, USA}

\date{\today}

\begin{abstract}
Superconducting quantum processors are a leading platform for implementing  practical quantum computation algorithms. Although superconducting quantum processors with hundreds of qubits have been demonstrated, their further scaling up is constrained by the physical size and cooling power of dilution refrigerators. This constraint can be overcome by constructing a quantum network to interconnect qubits hosted in different refrigerators, which requires microwave-to-optical transducers to enable low-loss signal transmission over long distances. Despite that various designs and demonstrations have achieved high-efficiency and low-added-noise transducers, a coherent photonic link between separate refrigerators has not yet been realized. In this work, we experimentally demonstrate coherent signal transfer between two superconducting circuits housed in separate dilution refrigerators, enabled by a pair of frequency-matched aluminum nitride electro-optic transducers connected via a 1-km telecom optical fiber. {With transducers at each node achieving >0.1\% efficiency, an overall 80 dB improvement in transduction efficiency over commercial electro-optic modulators is attainable, paving the way towards a fully quantum-enabled link.} This work provides critical design guidelines towards scalable superconducting quantum networks interconnected by photonic links.
\end{abstract}
\maketitle

\noindent 

\begin{figure}[t]
\capstart
\centering
\includegraphics[width=\linewidth]{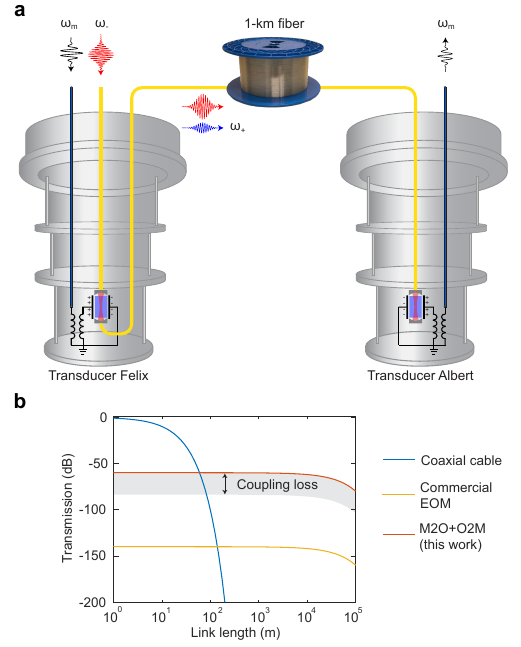}
\caption{\textbf{Photonic link between refrigerators.} (a) Schematic of the photonic link. Felix up-converts microwave photons ($\omega_{\text{m}}$) to optical photons ($\omega_{+}$) with a parametric pump light ($\omega_{-}$). The up-converted optical photons propagate through a 1-km fiber and are down-converted to microwave photons at Albert. (b) Link transmission for different methods. The transmission through a coaxial cable (blue line) exhibits 1~dB/m attenuation. Both commercial EOMs (yellow line) and our transducers (orange line) enable transmission through fibers with 0.2~dB/km attenuation. A pair of EOMs have a total transduction efficiency of $-140$~dB, while our transducer pair has $-60$~dB on-chip transduction efficiency. The shaded area indicates the fiber-to-chip coupling loss ($-23.7$~dB in total) and can be mitigated by optimizing grating couplers.}
\label{Fig:overview}
\end{figure}

Superconducting qubits have widely been considered as a promising route towards quantum computational advantage \cite{neill2018blueprint}{}. Recent advancements in superconducting quantum circuits have enabled the implementation of quantum processors with hundreds of qubits \cite{acharya2025quantum, gao2025establishing, boixo2014evidence}. However, several millions of qubits are needed to solve practical problems \cite{webber2022impact}, and the operation of such a large number of qubits leads to significantly higher wiring complexity and thermal load, which are constrained by the physical size and cooling power of a single dilution refrigerator \cite{krinner2019engineering}. A flexible and cost-effective solution is to build a quantum network by connecting qubits inside separate refrigerators. Connecting qubits with standard coaxial cables, however, is impractical due to the high microwave loss and thermal noise in room temperature environment. A meter-scale superconducting waveguide cooled to millikevin temperature between two refrigerators has been constructed to connect remote qubits \cite{magnard2020microwave, yam2025cryogenic}. However, such bulky waveguides cannot be easily extended to longer distances. Another approach is the photonic link through the use of microwave-to-optical (M2O) transducers. Conventional M2O transducers are commercially available lithium niobate electro-optic modulators (EOM)\cite{wooten2000review}{}. However, the transduction efficiency of a commercial EOM is approximately $-70$~dB, resulting in a typical half-wave voltage $V_{\pi} \approx 5 V$ that corresponds to watt-level microwave modulation power \cite{youssefi2021cryogenic,shen2024photonic}. By contrast, the control and readout of the qubit states are typically performed with picowatt-level microwave power.

\begin{figure*}[t]
\capstart
\centering
\includegraphics[width=\linewidth]{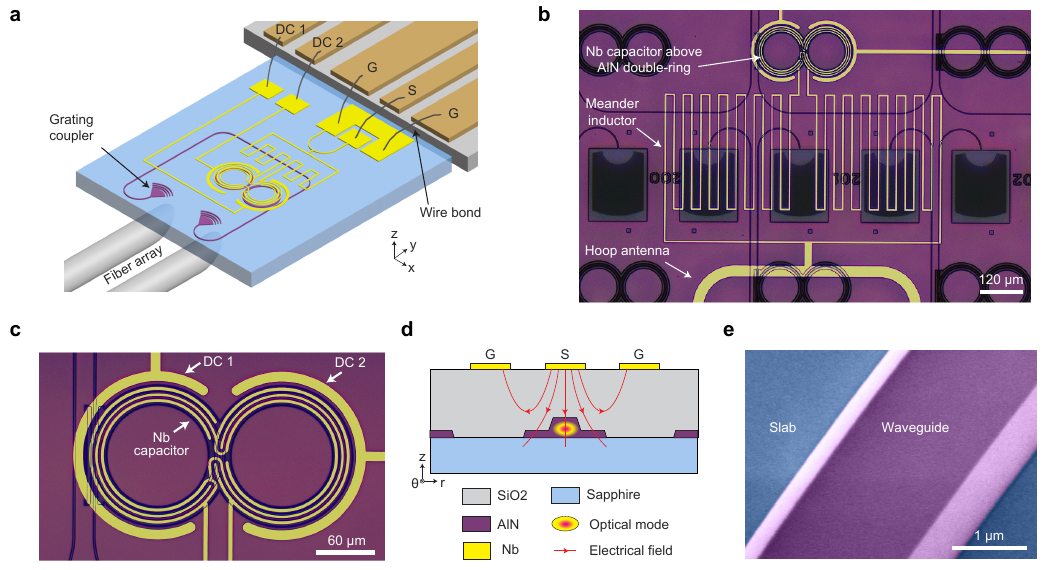}
\caption{\textbf{Design of electro-optic transducers.} (a) Schematic of the AlN electro-optic transducer. (b) Micrograph of the superconducting resonator. (c) Magnified view of the capacitor electrodes above the optical double-ring. (d) Cross-section view of the AlN ring and electrodes. (e) False-color scanning-electron microscopy image of the AlN waveguide before the cladding deposition.}
\label{Fig:scheme}
\end{figure*}

To enhance the transduction efficiency, many mechanisms have been experimentally explored, such as electro-optics \cite{fan2018superconducting, arnold2025all, warner2025coherent, xu2021bidirectional, mckenna2020cryogenic, pintus2022ultralow, delaney2022superconducting, holzgrafe2020cavity, rueda2016efficient}, magneto-optics \cite{pintus2022integrated, zhu2020waveguide, shen2022coherent, chai2022single}, electro-optomechanics \cite{mirhosseini2020superconducting, zhao2025quantum, van2025optical, andrews2014bidirectional, jiang2020efficient, han2020cavity, weaver2024integrated, zhou2024electrically, arnold2020converting, meesala2024non, meesala2024quantum}, and rare-earth ions \cite{xie2025scalable, rochman2023microwave, nicolas2023coherent, xie2021characterization}. Among various designs, transducers based on photonic integrated circuits are particularly attractive due to their small mode volume, enhanced optical nonlinearity, and compact footprint \cite{han2021microwave}. To enhance transduction efficiency, resonators are commonly employed to enhance the intra-cavity photon number at the cost of a smaller bandwidth. To date, more than 1\% on-chip transduction efficiency has been reported on integrated photonic platforms with low added noise \cite{warner2025coherent, zhao2025quantum, xie2025scalable}. For a triply resonant electro-optic transducer \cite{mckenna2020cryogenic}, the resonance frequencies of the microwave cavity ($\omega_{\text{m}}$) and the optical cavity (red sideband at $\omega_{-}$ and blue sideband at $\omega_{+}$) need to be precisely aligned to satisfy $\omega_{\text{m}} = \omega_{+}-\omega_{-}$, which we refer to as the intra-cavity frequency-matching condition. It is worth noting that resonance frequency tuning in a cryogenic environment is technically challenging in itself, due to the inapplicability of the commonly used thermal tuning method. In addition, the interconnection between two refrigerators requires two transducers with the same blue sideband frequency as $\omega_{+,1}=\omega_{+,2}$, which we refer to as the inter-cavity frequency-matching condition. Therefore, compared to a standalone transducer, more degrees of freedom to tune the resonance frequency are required for a transducer pair, which is the critical roadblock for fridge-to-fridge interconnection with cavity-based transducers. A recent breakthrough has demonstrated optical interference between a pair of transducers inside a single refrigerator based on the inherently matching frequencies of the atomic transitions in ytterbium ions \cite{xie2025scalable}. However, the ytterbium transition wavelength at 984.5~nm is not directly compatible with telecom fiber networks. Hence, the photonic link between dilution refrigerators over telecom fibers remains to be implemented.

In this work, we demonstrate a 1-km photonic link that connects superconducting circuits in two separate dilution refrigerators as shown in Fig.~\ref{Fig:overview}(a). Each transducer consists of a superconducting resonator and an optical resonator that are coupled via electro-optic effects \cite{boydNLO}. The first transducer, referred to as Felix, performs M2O frequency up-conversion. Microwave pulses are sent to Felix and up-converted to optical pulses. The optical pulses propagate through a 1-km telecom fiber and arrive at the second transducer, Albert, for optical-to-microwave (O2M) frequency down-conversion. Two transducers are designed to have slightly different optical free spectral ranges (FSRs), and the Vernier effect guarantees that we can find a resonance pair with a small frequency mismatch. Our design allows us to individually tune both the blue and red sideband frequency in each transducer, and thus both intra-cavity and inter-cavity frequency matching can be achieved simultaneously. Both transducers are characterized to have more than 0.1\% on-chip transduction efficiency. Thanks to the low propagation loss 0.2~dB/km in telecom fibers, $-60$~dB transmission can be available over a 1-km photonic link, as depicted in Fig.~\ref{Fig:overview}(b). As a comparison, electrical coaxial cables have a propagation loss of $\sim$1~dB/m at gigahertz frequency, and thus as high as $1000$~dB attenuation is expected for a 1-km coaxial link. By contrast, when using a pair of commercial EOMs, the total transmission is $-140$~dB due to the $-70$~dB transduction efficiency of each EOM. Therefore, our work shows up to 80~dB improvements compared to EOMs and  thus presents an important step towards photonic linked superconducting quantum networks. 

\section*{Results}
\noindent \textbf{Design of electro-optic transducers}

\noindent Figure~\ref{Fig:scheme}(a) shows the schematic of a packaged transducer. Grating couplers are glued at the back side of the chip for fiber-to-chip coupling \cite{zhou2025high}. An aluminum nitride (AlN) photonic molecule structure \cite{liao2020photonic} based on evanescently coupled double rings is used to generate optical resonance doublets \cite{fu2021cavity, liu2023aluminum} (see Methods). A niobium (Nb) superconducting resonator is deposited above the AlN double-ring as shown in Fig.~\ref{Fig:scheme}(b). Nb is a superconductor with negligible kinetic inductance and thus shows weaker resonance frequency drift under optical excitations compared to other high-kinetic-inductance superconductors such as niobium nitride \cite{li2025fast}. The Nb resonator consists of a capacitor and a meander inductor. The length of the meander inductor is adjusted to match the microwave resonance frequency to the optical doublet spectral gap. The magnified view of the capacitor is shown in Fig.~\ref{Fig:scheme}(c). The polarity of the resonator electrodes on two rings are reversed for the following reason. For a double-ring resonator, we denote the annihilation operators of the individual optical modes in each ring by $\hat{a}_{1}$ and $\hat{a}_{2}$. The evanescent coupling between two rings generates a hybridized blue sideband mode $\hat{a}_{+} = (\hat{a}_{1} + \hat{a}_{2})/\sqrt{2}$ and a red sideband mode $\hat{a}_{-}=(\hat{a}_{1} - \hat{a}_{2})/\sqrt{2}$ with eigenfrequencies $\omega_{+}$ and $\omega_{-}$ respectively \cite{xu2021bidirectional}. To satisfy the phase-matching condition between $\hat{a}_{+}$, $\hat{a}_{-}$ and the microwave mode $\hat{a}_\mathrm{m}$, the polarity of electrodes on two rings are reversed to introduce a $\pi$ phase, which is manifested by the capacitor charge distribution in Extended Data Fig.~\ref{Fig:charge}. The capacitor generates electric fields along the z axis that interact with the optical transverse-magnetic (TM) modes in AlN ring through the electro-optic coefficient $r_{33}\approx 1$~pm/V as depicted in Fig.~\ref{Fig:scheme}(d). Two additional electrodes are placed near the resonator electrodes to enable individual optical resonance frequency tuning of each optical ring by applying direct current (DC) voltages. The superconducting resonator is inductively coupled to a hoop antenna which is wire bonded to a printed circuit board (PCB). The entire resonator is connected to the ground of PCB via a nanowire between the hoop antenna and the meander inductor (see Fig.~\ref{Fig:scheme}(b)). The resonator electrodes above the AlN ring also act as the DC ground and thus generate DC electric fields to tune the refractive index of AlN. Hence, we can achieve individual DC tuning for two rings, while the resonator capacitor covers the entire double-ring for maximized transduction efficiency. Chlorine etching is used to etch the AlN waveguides with low roughness \cite{liu2015smooth}, and the AlN sidewall image is shown in Fig.~\ref{Fig:scheme}(e). More details of the device fabrication can be found in Methods and Extended Data Fig.~\ref{Fig:fabflow}.



\vspace{0.6cm}
\noindent \textbf{Frequency matching between the transducer pair}

\begin{figure*}[t]
\centering
\includegraphics[width=\linewidth]{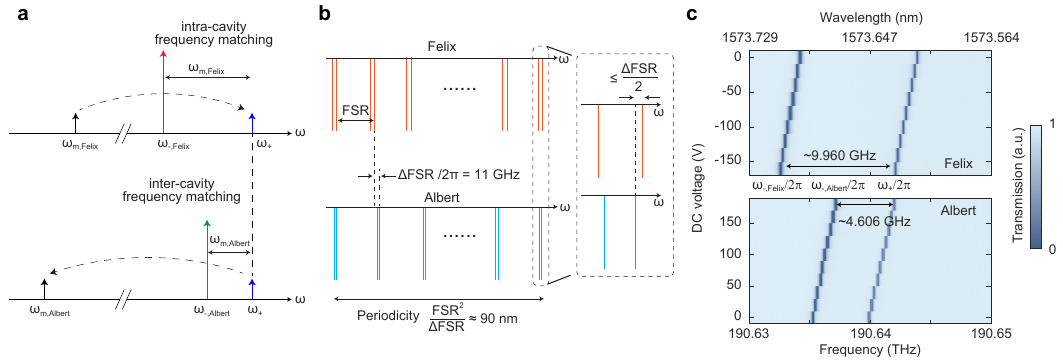}
\caption{\textbf{Frequency matching of the transducer pair.} (a) Frequency conversion diagram of the transducer pair. Felix up-converts microwave photons ($\omega_{\text{m,Felix}}$) to optical photons ($\omega_{+}$) with an optical pump at $\omega_{-,\text{Felix}}$. Albert down-converts optical photons ($\omega_{+}$) to microwave photons ($\omega_{\text{m,Albert}}$) with an optical pump at $\omega_{-,\text{Albert}}$. (b) Vernier effect for inter-cavity frequency matching. The top and bottom panels show the optical resonance spectrum of two transducers. The Vernier periodicity is $\text{FSR}^2 / \Delta \text{FSR} \approx 90$~nm, and the frequency mismatch is upper bounded to $\Delta \text{FSR}/2$ as indicated in the dashed box. (c) Optical resonance frequency tuning by applying DC voltages to both DC electrodes in each transducer.}
\label{Fig:resonance}
\end{figure*}

\noindent To interconnect a pair of transducers, both intra-cavity and inter-cavity frequency matching conditions need to be satisfied as shown in Fig.~\ref{Fig:resonance}(a). Due to the limited tuning range, it is necessary to fabricate two transducers with a small inter-cavity frequency mismatch. To achieve this, we adopt the form of asymmetric photonic molecules: the ring radius is 61.7~\textmu m for Felix and 60.0~\textmu m for Albert. Two transducers have an optical $\text{FSR}\approx 2\pi \cdot 353$~GHz (2.921~nm), and their FSR difference is $\Delta \text{FSR}=2\pi \cdot 11$~GHz. The optical resonance spectra for transducers are schematically illustrated in Fig.~\ref{Fig:resonance}(b). The Vernier effect ensures that we can always find a nearly frequency-matched resonance pair with a periodicity of $\text{FSR}^2/ \Delta \text{FSR}\approx 2\pi \cdot 11.3$~THz (90~nm). Therefore, inter-cavity frequency matching can be achieved in the range of $1550 \pm 45$~nm, well localized in the telecom band. In addition, as indicated in the dashed box in Fig.~\ref{Fig:resonance}(b), the minimum frequency mismatch is upper bounded to $\Delta \text{FSR}/2=2\pi \cdot 5.5$~GHz. By using two DC electrodes in each transducer, we are able to tune $\omega_+$ and $\omega_-$ simultaneously, and thus both intra-cavity frequency matching $\omega_\mathrm{m}=\omega_+-\omega_-$ and inter-cavity frequency matching $\omega_{+,\text{Felix}} = \omega_{+,\text{Albert}}$ can be achieved. In the absence of DC tuning voltage, we have $\omega_{+,\text{Albert}}= 2\pi \cdot 190.6400~\text{THz}$ and $\omega_{+,\text{Felix}}= 2\pi \cdot 190.6438~\text{THz}$, and thus the frequency mismatch is as low as 3.8~GHz. By applying 160~V to Albert and $-160$~V to Felix, we are able to obtain $\omega_{+,\text{Felix}} = \omega_{+,\text{Albert}}$. By further adjusting the two DC tuning voltages in each transducer, both intra-cavity and inter-cavity frequency matching can be realized near 1573.6~nm as shown in Fig.~\ref{Fig:resonance}(c).

\vspace{0.6cm}
\noindent \textbf{Characterization of transducers}

\begin{figure*}[t]
\capstart
\centering
\includegraphics[width=\linewidth]{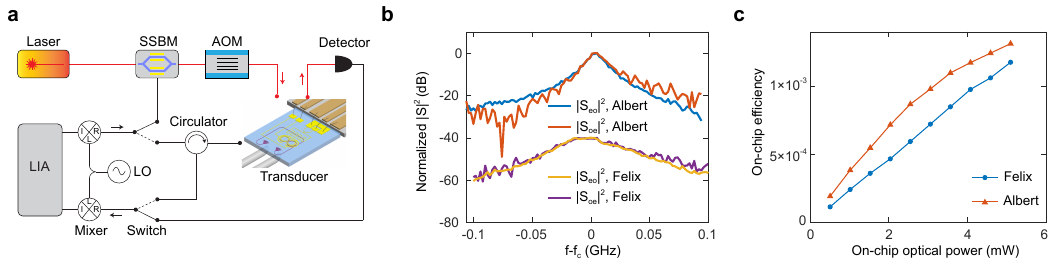}
\caption{\textbf{Transduction efficiency characterization.} (a) Simplified schematic of the setup to characterize the transduction efficiency. SSBM, single sideband modulator; AOM, acousto-optic modulator; LO, local oscillator. (b) Experimentally measured, normalized M2O and O2M scattering matrix spectra. The spectra for Felix is offset by 40~dB. The center frequency $f_c$ for Albert and Felix is 4.606~GHz and 9.960~GHz, respectively. (c) On-chip transduction efficiency at different on-chip optical peak powers.
}
\label{Fig:Seo}
\end{figure*}

\noindent To calibrate the on-chip transduction efficiency, we use the setup in Fig.~\ref{Fig:Seo}(a) to measure the scattering matrix spectra of microwave reflection $S_{\text{ee}}$, optical reflection $S_{\text{oo}}$, M2O conversion $S_{\text{oe}}$, and O2M conversion $S_{\text{eo}}$. The transducer is mounted in a dilution refrigerator anchored at 50~mK. An optical single-sideband modulator (SSBM) is used to generate a frequency-tunable optical sideband. The laser is modulated by an acousto-optic modulator (AOM) to generate optical pulses. We use a pulse width of 2.5~\textmu s and a repetition rate of 1~kHz to suppress the light-induced heating effect. The optical pulses are coupled to the TM mode in the AlN waveguide with an insertion loss of $-5.93$~dB per coupler. During the measurement, the laser frequency is aligned to the red sideband mode at $\omega_-$ as the parametric pump. The output microwave signal synthesized from a lock-in amplifier (LIA) is sent to either the microwave port of the transducer or the SSBM by using a microwave switch. The microwave signal from either the optical detector or the superconducting resonator is sampled by the LIA. By adjusting the configuration of two microwave switches, we can measure $S_{\text{ee}}$, $S_{\text{oo}}$, $S_{\text{oe}}$, and $S_{\text{eo}}$ without changing any optical or microwave wiring. The on-chip transduction efficiency can thus be experimentally obtained as \cite{xu2021bidirectional, andrews2014bidirectional}
\begin{equation}
\eta = \frac{S_\text{oe,pk} S_\text{eo,pk} }{S_\text{oo,bg} S_\text{ee,bg}},
\label{eq:eta}
\end{equation}
where $S_\text{oe,pk}$ and $S_\text{eo,pk}$ are the on-resonance peak value of the M2O and O2M conversion spectra, and $S_\text{oo,bg}$ and $S_\text{ee,bg}$ are the off-resonance background value of the optical and microwave reflection spectra, respectively. The measured normalized conversion spectra for both Felix and Albert are presented in Fig.~\ref{Fig:Seo}(b). The 3~dB conversion bandwidth is 34.5~MHz for Felix and 13.0~MHz for Albert, which agrees with the microwave mode total loss rate $\kappa_{\text{m,Felix}}= 2\pi \cdot 38.0$~MHz and $\kappa_{\text{m,Albert}}=2\pi \cdot 13.9$~MHz (see Extended Data Table~\ref{table:parameter}). The measured on-chip transduction efficiency is shown in Fig.~\ref{Fig:Seo}(c). Both transducers can achieve 0.1\% transduction efficiency at an optical power near 4~mW. At a higher optical power, the efficiency starts to show saturation effect, which is attributed to the reduced microwave quality factor upon optical excitations \cite{xu2021bidirectional}. The electro-optic single-photon coupling rate for both transducers is characterized to be $g_{\text{eo}} \approx 2\pi\cdot 280$~Hz, as discussed in Methods and Extended Data Table~\ref{table:parameter}. The full schematic of the measurement setup is presented in Extended Data Fig.~\ref{Fig:conversion_setup}.

\begin{figure}[b]
\capstart
\centering
\includegraphics[width=\linewidth]{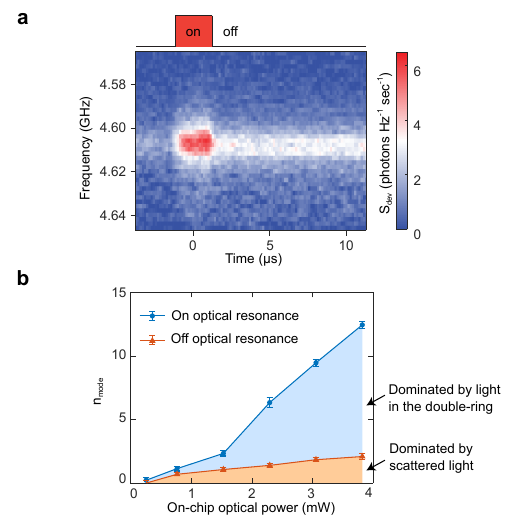}
\caption{\textbf{Characterization of added noise.} (a) Time-resolved added noise spectrum $S_{\text{dev}}$ with an on-chip optical power of 1.5~mW. (b) Microwave mode thermal occupancy $\bar{n}_{\text{mode}}$ at different optical powers under on-resonance and off-resonance optical excitations. The error bar represents one standard error of the mean.
}
\label{Fig:noise}
\end{figure}

\begin{figure*}[t]
\capstart
\centering
\includegraphics[width=\linewidth]{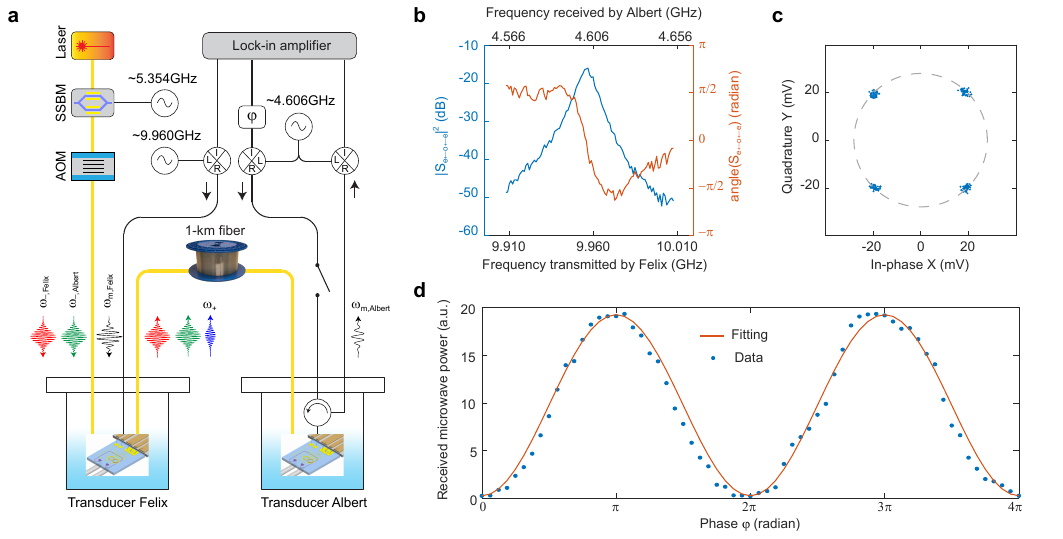}
\caption{\textbf{Characterization of the photonic link.} (a) Simplified schematic of the photonic link. By using a laser and an optical SSBM, two optical pumps at $\omega_{-,\text{Felix}}$ and $\omega_{-,\text{Albert}}$ are generated and co-propagate in fiber. A microwave pulse ($\omega_{\text{m,Felix}}$) is up-converted to an optical pulse ($\omega_{+}$) at Felix, which propagates through a 1-km fiber spool and is down-converted to a microwave pulse ($\omega_{\text{m,Albert}}$) at Albert. (b) Measured amplitude and phase of the M2O2M scattering matrix spectrum $S_{e \leftarrow o \leftarrow e}$. (c) Measured quadrature distributions over the photonic link. Each quadrature measurement contains 50 data points. (d) Received microwave power at different local oscillator phases. The microwave switch is enabled for interference measurement.
}
\label{Fig:fridge}
\end{figure*}

We next characterize the added noise of our transducer. A 30~dB attenuator is mounted on a variable temperature stage (VTS) at the mixing chamber as a controllable blackbody radiation source. The output of the VTS is sent to a transducer, and the reflected microwave signal is amplified by a traveling-wave parametric amplifier (TWPA) \cite{macklin2015near} at 50~mK and subsequently a high-electron-mobility transistor (HEMT) amplifier at 4~K. Due to the bandwidth limitation of our TWPA, we only characterize the added noise of Albert. The output signal from the refrigerator is further amplified by room temperature amplifiers and then detected by a spectrum analyzer. The spectrum analyzer is implemented by a LIA that allows us to measure the time-resolved power spectrum. The details of the setup are presented in Methods and Extended Data Fig.~\ref{Fig:twofridgesetup}. Following the procedure in refs. \cite{fu2021cavity, xu2020radiative}, we first calibrate the gain and added noise of the output amplification chain by sweeping the VTS temperature and measure the corresponding output power spectrum. We then send optical pulses to the transducer and measure the corresponding time-resolved power spectrum (see Methods for details). The output noise power spectral density can be expressed as \cite{fu2021cavity}
\begin{equation}
S_{\text{dev}}(\omega) = R(\omega) \bar{n}_{\text{ex}} + [ 1-R(\omega)] \bar{n}_{\text{en}} + \Delta \bar{n}_{\text{out, add}},
\label{eq:Sdev}
\end{equation}
where $R(\omega)$ is the power reflection spectrum of the superconducting resonator, $\bar{n}_{\text{ex}}$ and $\bar{n}_{\text{en}}$ are the thermal bath occupancy of the intrinsic and external bath respectively, and $ \Delta \bar{n}_{\text{out, add}} $ is additional output line added noise induced by light. The microwave mode thermal occupancy can thus be calculated as \cite{ xu2020radiative}
\begin{equation}
\bar{n}_{\text{mode}} = \frac{ \kappa_{\text{m,in}} \bar{n}_{\text{en}}  + \kappa_{\text{m,ex}} \bar{n}_{\text{ex}} }{  \kappa_{\text{m,in}} + \kappa_{\text{m,ex}}},
\label{eq:nmode}
\end{equation}
where $\kappa_{\text{m,in}}$ and $\kappa_{\text{m,ex}}$ are the intrinsic and external loss rate of the microwave mode. The measured time-resolved $S_{\text{dev}}$ is presented in Fig.~\ref{Fig:noise}(a) with an on-chip optical power of 1.5~mW. This noise is nearly 10~dB higher than our previous result \cite{fu2021cavity}, and we attribute it to the small gap size of $\sim 2.5$~\textmu m between the Nb electrodes and the AlN ring, which is much smaller than the 14~\textmu m gap size in our previous work\cite{fu2021cavity}. This is also evidenced by the measured $\bar{n}_{\text{mode}}$ in Fig.~\ref{Fig:noise}(b), where we present the results under on-resonance and off-resonance optical excitations. For on-resonance excitations, the noise is mainly induced by the light inside the AlN ring, while for off-resonance excitations, the noise is mainly induced by the scattered light from grating couplers. It can be seen that on-resonance $\bar{n}_{\text{mode}}$ is significantly higher than off-resonance $\bar{n}_{\text{mode}}$, suggesting that the light inside the AlN ring is responsible for the noise. This result is distinct from our previous results, where the noise photon is nearly unchanged for both on-resonance and off-resonance excitations. Hence, we believe that the noise photon can be significantly suppressed by appropriately increasing the gap size between the electrodes and AlN ring. Although $g_{\text{eo}}$ will be reduced with an increased gap size, the microwave quality factor can improve due to reduced piezoelectric loss, and thus lower noise may be obtained without compromising the transduction efficiency, which we leave for future study.

\vspace{0.6cm}
\noindent \textbf{Photonic link between two refrigerators}

\noindent The simplified schematic for the photonic link between two refrigerators is presented in Fig.~\ref{Fig:fridge}(a). We first apply appropriate DC voltages such that the intra-cavity and inter-cavity frequency-matching conditions are satisfied. The laser frequency is tuned to $\omega_{-,\text{Felix}}$ as the pump for M2O transduction at Felix. We then use an optical SSBM to generate a sideband at $\omega_{-,\text{Albert}}$ as the pump for O2M transduction at Albert. The microwave pulses from the LIA are frequency up-converted to $\omega_{\text{m,Felix}}\approx 2\pi \cdot 9.960$~GHz and sent to Felix. Felix performs M2O transduction and up-convert microwave pulses to optical pulses at $\omega_+$. The optical pulses propagate through a 1-km single-mode fiber spool (Corning SMF-28e) and arrive at Albert. The optical pulses at frequency $\omega_{-,\text{Albert}}$ work as the pump for O2M transduction to down-convert the photons at $\omega_+$ to microwave photons at $\omega_{\text{m,Albert}} \approx 2\pi \cdot 4.606$~GHz. The on-chip optical pump power for both transducers is $\sim 3$~mW. The detailed schematic is discussed in Methods and Extended Data Fig.~\ref{Fig:twofridgesetup}.

We first characterize the microwave-to-optical-to-microwave (M2O2M) fridge-to-fridge scattering matrix $S_{\text{e} \leftarrow \text{o} \leftarrow \text{e}}$ of the photonic link, and the result is shown in Fig.~\ref{Fig:fridge}(b).
The well-defined phase profile of the scattering matrix validates the phase coherence of transducers. We then use the photonic link to perform data transfer by employing quadrature phase-shift keying (QPSK). We keep the amplitude unchanged and only tune the phases of microwave pulses at Felix. The measured constellation diagram is shown in  Fig.~\ref{Fig:fridge}(c), where each quadrature measurement contains 50 data points. Compared to commercial EOMs requiring up to 30~dBm microwave modulation power, the on-chip microwave power we apply at Felix is less than $-30$~dBm. The clean constellation diagram suggests a low bit error rate, and thus the microwave power can be further reduced while maintaining a reasonable signal-to-noise ratio. To further test the coherence of transducers, we measure the interference between the optically generated microwave pulses and a local oscillator at Albert. We measure the output microwave power while sweeping the phase of the local oscillator, and the measured interference patten is shown in Fig.~\ref{Fig:fridge}(d). The interference pattern overlaps well with a sine function and thus validates the coherence of the 1-km photonic link.

\section*{Discussion}

\noindent In this work, we have demonstrated a 1-km photonic link that connects superconducting resonators in two refrigerators. By adopting the form of asymmetric photonic molecules with slightly different FSRs, we can always find a resonance pair with sufficiently small frequency mismatch in the telecom band. Our AlN transducers allow for more than 0.1\% transduction efficiency, and thus the total link transmission in principle exceeds that of conventional EOMs by up to 80~dB. Higher transduction efficiency can be available by further reducing mode volumes and improving optical and microwave quality factors. The phase coherence of the photonic link is validated by measuring the M2O2M scattering matrix as well as the interference between the local oscillator and the optically generated microwave photons. {In addition, our transducer pair is readily applicable to build a quantum-enabled link by generating remotely entangled microwave photons between two refrigerators using heralding-based schemes \cite{duan2001long, zhong2020proposal, rueda2019electro}, and thus the 50\% efficiency threshold\cite{wolf2007quantum} can be bypassed.} We believe that our work presents an important step towards large-scale superconducting quantum networks.

\def\bibsection{\section*{References}}

\clearpage

\section*{{\normalsize{}Methods}}

\noindent \textbf{Photonic molecules transduction efficiency}

\noindent The photonic molecule structure in our work is realized by evanescently coupling two identical microring resonators. We denote the annihilation operators of the individual optical modes in ring 1 and ring 2 by $\hat{a}_{1}$ and $\hat{a}_{2}$ respectively, and the corresponding electric field spatial profiles are denoted as $\mathbf{u}_1(\mathbf{r})$ and $\mathbf{u}_2(\mathbf{r})$, where $\mathbf{r}$ is the coordinate vector. The capacitor of a superconducting resonator is deposited above the photonic molecule, and the electric field of the microwave mode $\hat{a}_{\text{m}}$ spans both microrings and is denoted by $\mathbf{u}_{\text{m}}(\mathbf{r})$. The electric field operator can thus be expressed as\cite{gerry2023introductory}
\begin{align}
\hat{\mathbf{E}}_n(\mathbf{r}) &= C_n \mathbf{u}_n (\mathbf{r})\hat{a}_n + h.c., \\
C_n &= \sqrt{ \frac{\hslash \omega_n}{ 2 \epsilon_0} \frac{1}{ V_{\text{eff},n}}  } ,\\
V_{\text{eff},n} &= \int_{V_{n}} dV \sum_{ij} \epsilon_{n,ij} u_{n,i}^{*} u_{n,j},
\end{align}
where $n = \{ 1,2,\text{m} \}$, $C_n $ is the normalization factor\cite{mckenna2020cryogenic}, $ V_{\text{eff},n}$ is the effective mode volume for mode $n$, $\omega_{n}$ is the eigenfrequency of each mode, $\epsilon_0$ is the vacuum permittivity, $\epsilon_{n,ij}$ is the relative permittivity matrix element of mode $n$, $V_1$ ($V_2$) is the integration space for ring 1 (ring 2), $V_{\text{m}}=V_1+V_2$ because the microwave capacitor spans both rings, and $u_{n,i}$ is the component of the spatial profile of mode $n$ along the $i$ axis. In our work, because we use optical TM modes for transduction and because two rings are identically designed, thus $\mathbf{u}_n(\mathbf{r}) = \hat{\mathbf{z}} \: u_{\text{o,z}}(r_n,z_n)e^{im\theta_n}$, where $m$ is the azimuthal mode number, $\hat{\mathbf{z}}$ is the unit vector along the z axis, $u_{\text{o,z}}$ is transverse optical mode profile along the z axis (see Fig.~\ref{Fig:scheme}(d)) for both rings, $\mathbf{r}_n=(r_n,\theta_n,z_n)$ is the local cylindrical coordinate for ring $n$ ($n=\{1,2\}$). It is worth noting that $u_{\text{o,z}}(r_n,z_n)$ is well confined in the AlN ring region $r_n\approx R$, with $R$ being the ring radius, and thus $\int_{V_n} dV=\iiint rdrdzd\theta\approx 2\pi R \iint drdz$, and the effective mode volume for both optical modes can be written as 
\begin{align}
\begin{split}
V_{\text{eff,o}}=2\pi R \iint  \epsilon_{\text{o,zz}}|u_{\text{o,z}}|^2 drdz,
\end{split}
\end{align}
where $\epsilon_{\text{o,zz}}$ is the relative permittivity for z-polarized optical modes. Consequently, we have 
\begin{align}
C_{\text{o}} =C_1=C_2= \sqrt{\frac{1}{2\pi R}} \sqrt{ \frac{\hslash \omega_{\text{o}}}{2\epsilon_0 }} \sqrt{\frac{1}{\iint  \epsilon_{\text{o,zz}}|u_{\text{o,z}}|^2 drdz}}
\end{align}
The transverse profile for microwave modes, as depicted by the red lines in Fig.~\ref{Fig:scheme}(d), can be expressed as $\mathbf{u}_{\text{m,t}}(\mathbf{r}) = u_{\text{m,r}}(r,z)\hat{\mathbf{r}} + u_{\text{m,z}}(r,z)\hat{\mathbf{z}}$. Because the superconducting resonator size is significantly smaller than the microwave wavelength, the microwave field is assumed to be spatially uniform along the ring and has no $\theta$ dependence, and thus we can write $\mathbf{u}_{\text{m}}(\mathbf{r}) = \mathbf{u}_{\text{m,t}}(\mathbf{r}_1) - \mathbf{u}_{\text{m,t}}(\mathbf{r}_2)$. The negative sign is caused by the reversed polarity of the electrodes above the two rings (see Fig.~\ref{Fig:scheme}(c) for the electrode geometry). Therefore, the effective microwave mode volume becomes 
\begin{align}
V_{\text{eff,m}}=4\pi R \iint  (\epsilon_{\text{m,zz}}|u_{\text{m,z}}|^2 + \epsilon_{\text{m,rr}}|u_{\text{m,r}}|^2) drdz,
\end{align}
where $4\pi R$ comes from the spatial integration over two rings. The system Hamiltonian can be written as \cite{mckenna2020cryogenic}
\begin{align}
\begin{split}
H = & \hslash \omega_1 \hat{a}_{1}^{\dagger} \hat{a}_{1}  + \hslash \omega_2 \hat{a}_{2}^{\dagger} \hat{a}_{2}^{} + \hslash \omega_{\text{m}} \hat{a}_{\text{m}}^{\dagger} \hat{a}_{\text{m}} \\
& + \hslash g_{\text{c}} (\hat{a}_{1}^{\dagger} \hat{a}_{2} + \hat{a}_{1} \hat{a}_{2}^{\dagger}) + H_{\text{eo}},
\end{split}
\end{align}
where $g_{\text{c}}$ is the evanescent coupling strength between $\hat{a}_{1}$ and $\hat{a}_{2}$, and $H_{\text{eo}}$ is the electro-optic interaction Hamiltonian to be discussed later. The evanescent coupling introduces the hybridized optical modes as
\begin{align}
\begin{split}
\hat{a}_{+}&=\cos \theta \hat{a}_1 + \sin \theta \hat{a}_2, \\
\hat{a}_{-}&=-\sin \theta \hat{a}_1 + \cos \theta \hat{a}_2,
\end{split}
\end{align}
where $\tan 2\theta = \frac{2g_{\text{c}}}{\omega_1 - \omega_2}$. The eigenfrequencies of the hybridized modes are 
\begin{align}
\begin{split}
\omega_{+}&= \frac{\omega_1+\omega_2}{2}+\sqrt{g_{\text{c}}^2+\left( \frac{\omega_1-\omega_2}{2} \right)^2}, \\
\omega_{-}&= \frac{\omega_1+\omega_2}{2}-\sqrt{g_{\text{c}}^2+\left( \frac{\omega_1-\omega_2}{2} \right)^2}.
\end{split}
\end{align}
In our design, the eigenfrequencies of two rings are identical $\omega_1=\omega_2$, and thus we have $\theta=\pi/4$, $\hat{a}_{+}=\frac{1}{\sqrt{2}}(\hat{a}_1 + \hat{a}_2)$ as the blue sideband mode, $\hat{a}_{-}=\frac{1}{\sqrt{2}}(\hat{a}_2 - \hat{a}_1)$ as the red sideband mode, $\omega_+=\omega_1+g_{\text{c}}$, and $\omega_-=\omega_1-g_{\text{c}}$. In our work, we adjust the design of the superconducting resonator to satisfy the condition $\omega_{\text{m}} = \omega_+ - \omega_-$. In the hybridized mode basis, the Hamiltonian can be rewritten as 
\begin{align}
H = & \hslash \omega_- \hat{a}_{-}^{\dagger} \hat{a}_{-}  + \hslash \omega_+ \hat{a}_{+}^{\dagger} \hat{a}_{+}^{} + \hslash \omega_{\text{m}} \hat{a}_{\text{m}}^{\dagger} \hat{a}_{\text{m}} + H_{\text{eo}}
\end{align}
We next present the expression for $H_{\text{eo}}$. The presence of the microwave field $\hat{\mathbf{E}}_{\text{m}}$ modifies the inverse permittivity matrix $\eta_{ij}$ for optical modes through the relation \cite{boydNLO} $\eta_{ij} = \eta_{ij}^{(0)} + \Delta \eta_{ij} $, where $\eta_{ij}^{(0)}$ is the linear inverse permittivity matrix in the absence of microwave fields, $\Delta \eta_{ij} = r_{ijk} \hat{E}_{\text{m},k}$ is the microwave-induced inverse permittivity change, $\hat{E}_{\text{m},k}$ is the microwave electric field operator along the $k$ axis, and $r_{ijk}$ is the electro-optic Pockels coefficient tensor. The electro-optic interaction Hamiltonian can be written as
\begin{align}
\begin{split}
H_{\text{eo}} &= \frac{1}{2 \epsilon_0}  \sum_{n=1}^{2} \int_{V_n} dV \sum_{ij} \Delta \eta_{ij} \hat{D}_{n,i} \hat{D}_{n,j} \\
& = \frac{ \epsilon_0}{2 } \sum_{n=1}^{2} \int_{V_n} dV  \sum_{ij}  r_{ijk}   \epsilon_{nii} \epsilon_{njj}  \hat{E}_{\text{m},k} \hat{E}_{n,i} \hat{E}_{n,j},
\end{split}
\end{align}
where $\hat{D}_{n,i}$ ($\hat{E}_{n,i}$) is the optical displacement field (electrical field) operator component along the $i$ axis in ring $n$ ($n=\{1,2\}$). The above equation can be significantly simplified as follows. Our design uses optical TM modes that have only the z-polarized component. In addition, our capacitor electrodes are designed to generate microwave fields along the z axis in the AlN ring to utilize the $r_{\text{zzz}} =r_{33}\approx1$~pm/V coefficient. Therefore, the interaction Hamiltonian can be simplified as
\begin{align}
\begin{split}
H_{\text{eo}} = \frac{ \epsilon_0 r_{33}   \epsilon_{\text{o,zz}}^2}{2 } \sum_{n=1}^{2} \int_{V_n} dV      \hat{E}_{\text{m},z} \hat{E}_{n,z} \hat{E}_{n,z}
\end{split}
\end{align}
Because $\hat{E}_{n,\text{z}} (\mathbf{r}) = C_{\text{o}}  u_{\text{o,z}} (\mathbf{r}_n) e^{im\theta_n} \hat{a}_n + h.c.  $ and $\hat{E}_{\text{m,z}} (\mathbf{r}) = C_{\text{m}} [ u_{\text{m,z}}(\mathbf{r}_1) - u_{\text{m,z}}(\mathbf{r}_2) ]\hat{a}_{\text{m}} + h.c. $, by neglecting the counter-rotating terms, the above Hamiltonian can be written as
\begin{align}
\begin{split}
H_{\text{eo}} = &\frac{ \epsilon_0 r_{33}   \epsilon_{\text{o,zz}}^2 C_{\text{m}}C_{\text{o}}^2 }{2 } \int_{V_1+V_2} dV  u_{\text{m,z}} |u_{\text{o,z}}|^2  \\
& \cdot (\hat{a}_{\text{m}} + \hat{a}_{\text{m}}^{\dagger} )   (\hat{a}_{1}  \hat{a}_{1}^{\dagger} + \hat{a}_{1}^{\dagger} \hat{a}_{1}  - \hat{a}_{2}  \hat{a}_{2}^{\dagger} - \hat{a}_{2}^{\dagger} \hat{a}_{2}   )
\end{split}
\end{align}
By using $\hat{a}_{1}=\frac{1}{\sqrt{2}}(\hat{a}_{+} -\hat{a}_{-})$, $\hat{a}_{2}=\frac{1}{\sqrt{2}}(\hat{a}_{+} +\hat{a}_{-})$ and further neglecting counter-rotating terms, the above equation can be simplified as
\begin{align}
\begin{split}
H_{\text{eo}} = &\frac{ \epsilon_0 r_{33}   \epsilon_{\text{ozz}}^2 C_{\text{m}}C_{\text{o}}^2 }{2 } \iint drdz  \: 4\pi R u_{\text{mz}} |u_{\text{oz}}|^2  \\
& \cdot -2(   \hat{a}_{\text{m}}  \hat{a}_{-}  \hat{a}_{+}^{\dagger}   + \hat{a}_{\text{m}}^{\dagger}  \hat{a}_{-}^{\dagger}  \hat{a}_{+} )
\end{split}
\end{align}
Therefore, the system Hamiltonian can be rewritten as 
\begin{align}
\begin{split}
H = & \hslash \omega_- \hat{a}_{-}^{\dagger} \hat{a}_{-}  + \hslash \omega_+ \hat{a}_{+}^{\dagger} \hat{a}_{+}^{} + \hslash \omega_{\text{m}} \hat{a}_{\text{m}}^{\dagger} \hat{a}_{\text{m}} \\
&+ \hslash g_{\text{eo}}(   \hat{a}_{\text{m}}  \hat{a}_{-}  \hat{a}_{+}^{\dagger}   + \hat{a}_{\text{m}}^{\dagger}  \hat{a}_{-}^{\dagger}  \hat{a}_{+} ),
\end{split}
\end{align}
where $g_{\text{eo}}$ is the electro-optic single-photon coupling rate, which can be explicitly expressed as
\begin{align}
\begin{split}
g_{\text{eo}} = & \sqrt{  \frac{ \hslash \omega_{\text{o}}  \omega_{\text{o}}   \omega_{\text{m}}   }{8 \pi \epsilon_0 R}} \frac{\iint drdz \: \epsilon_{\text{o,zz}}^2 r_{33}  |u_{\text{o,z}}|^2 u_{\text{m,z}}    }{\iint drdz \epsilon_{\text{o,zz}} |u_{\text{o,z}}|^2   } \\
& \cdot \frac{1}{\sqrt{\iint  (\epsilon_{\text{m,rr}} |u_{\text{m,r}}|^2 + \epsilon_{\text{m,zz}} |u_{\text{m,z}}|^2) drdz }}
\end{split}
\end{align}
The transverse spatial profiles in the above equation $u_{\text{m,r}}$, $u_{\text{m,z}}$, and $u_{\text{o,z}}$ can be obtained by using a numerical mode solver. We use COMSOL Multiphysics to solve the transverse spatial profiles, and the computed result is $g_{\text{eo}}=2\pi\cdot 264$~Hz, which is close to the experimental value $g_{\text{eo}}\approx 2\pi\cdot 280$~Hz presented in Extended Data Table.~\ref{table:parameter}.

When a pump light is tuned to $\omega_-$, the intra-cavity pump photon number is $n_{-}= \frac{P}{ \hslash \omega_{-}} \frac{4 \kappa_{-,\text{ex}}}{\kappa_{-}^{2}}$. The $\hat{a}_-$ operator can be treated as a classical number with magnitude $a_-=\sqrt{n_-}$, and the system Hamiltonian be further simplify to 
\begin{align}
\begin{split}
H = &  \hslash \omega_+ \hat{a}_{+}^{\dagger} \hat{a}_{+}^{} + \hslash \omega_{\text{m}} \hat{a}_{\text{m}}^{\dagger} \hat{a}_{\text{m}} + \hslash G_{\text{eo}}(   \hat{a}_{\text{m}}   \hat{a}_{+}^{\dagger}   + \hat{a}_{\text{m}}^{\dagger}   \hat{a}_{+} ),
\end{split}
\end{align}
where $G_{\text{eo}} = \sqrt{n_-} g_{\text{eo}}$ is the enhanced electro-optic coupling rate. Following ref.\cite{fan2018superconducting}, the cooperativity is $C=\frac{4G_{\text{eo}}^2}{\kappa_{\text{m}} \kappa_{+}}$, and the on-chip transduction efficiency can be expressed as 
\begin{equation}
\eta_{}= \frac{\kappa_{+,\text{ex}} }{ \kappa_{+} }  \frac{ \kappa_{\text{m,ex}} }{ \kappa_{\text{m}}  } \frac{4C}{(1+C)^2}
\label{eq:efficiency}
\end{equation}
Here $\kappa_{-,\text{ex}}$, $\kappa_{+,\text{ex}}$, $\kappa_{\text{m,ex}}$ are the external coupling rate of the optical red sideband mode, the optical blue sideband mode, and the microwave mode, respectively. $\kappa_{-}$, $\kappa_{+}$, $\kappa_{\text{m}}$ are the corresponding total coupling rate and are defined as $\kappa_{-}=\kappa_{-,\text{ex}} + \kappa_{-,\text{in}}$, $\kappa_{+}=\kappa_{+,\text{ex}} + \kappa_{+,\text{in}}$, and $\kappa_{\text{m}}=\kappa_{\text{m,ex}} + \kappa_{\text{m,in}}$, where $\kappa_{-,\text{in}}$, $\kappa_{+,\text{in}}$, and $\kappa_{\text{m,in}}$ are the intrinsic coupling rate of the optical red sideband mode, the optical blue sideband mode, and the microwave mode, respectively.

\vspace{0.4cm}
\noindent \textbf{Device fabrication}

\noindent The fabrication process flow is displayed in Extended Data Fig.~\ref{Fig:fabflow}. A 1~\textmu m single-crystalline AlN thin film is deposited on a 430~\textmu m thick double-side-polished sapphire substrate by metal-organic chemical vapor deposition. We deposit silicon dioxide (SiO2) as a hard mask by plasma-enhanced chemical vapor deposition (PECVD). The negative-tone electron-beam (ebeam) resist (CSAR 62) is patterned by EBPG 5200 to etch the SiO2 hard mask by CHF3/O2, and 400~nm AlN is subsequently etched by Cl2/BCl3/Ar. The remaining SiO2 hard mask is removed by dipping in buffered oxide etch. We then use PECVD to deposit SiO2 as the hard mask for second round of patterning. We use positive-tone ebeam resist (Ma-N 2405) to etch the SiO2 hard mask by CHF3/O2, and the remaining 600~nm AlN is etched by Cl2/BCl3/Ar. We deposit 2.5~\textmu m SiO2 as the cladding by PECVD and anneal the chip at 1000~\textdegree C for 2 hours. We then spin coat photoresist on the chip and expose the resist by a maskless aligner (Heidelberg MLA150). A layer of 100~nm Nb is deposited by ebeam evaporation in an ultra-high-vacuum environment ($\sim 10^{-9}$ torr) as the superconducting resonator. The resist is lifted off by soaking the chip in N-Methylpyrrolidone (NMP). The chip is cleaved by using an ultraviolet laser cutter. The chip is then glued to a cleaved sapphire substrate, and the sapphire substrate is glued to a customized copper plate which has a hole at its edge to allow fiber arrays to access the back side of the chip. Aluminum wires are used to connect the on-chip electrodes to PCBs by using a wire bonder. A fiber array is glued to the chip back side by using ultraviolet-curable epoxy. A copper cover is then applied to enclose the entire chip to block undesirable radiations in the refrigerator.

\vspace{0.4cm}
\noindent \textbf{Transduction efficiency characterization}

\noindent The experimental setup for transduction efficiency characterization is shown in Extended Data Fig.~\ref{Fig:conversion_setup}. AOMs and erbium-doped fiber amplifiers (EDFAs) are used to generate optical pulses with sufficient peak power.  An optical SSBM is used to generate a frequency-tunable optical sideband. The pulse duration is 2.5~\textmu s, and the period is 1~ms. Mixers and a signal generator are used for microwave frequency up and down conversion between gigahertz-level radio-frequency (RF) signals for transducers and the megahertz-level intermediate-frequency signals from the LIA. Two $1 \times 2$ microwave switches are used to measure $S_{\text{oo}}$, $S_{\text{ee}}$, $S_{\text{oe}}$, $S_{\text{eo}}$ individually without changing any optical and microwave wiring. For $S_{\text{oo}}$ measurement, LIA output signals are directed to the optical SSBM, and the signals generated by the optical detector are measured by the LIA. For $S_{\text{ee}}$ measurement, LIA output signals are directed to the transducer, and the reflected signals are measured by the LIA. For $S_{\text{eo}}$ measurement, LIA output signals are directed to the SSBM, and the microwave signals from the transducer are measured by the LIA. For $S_{\text{oe}}$ measurement, LIA output signals are sent to the transducer, and the signals from the optical detector are measured by the LIA. During the measurements, the laser frequency is fixed at the pump frequency $\omega_-$ of the transducer, and appropriate DC voltages are applied to realize frequency matching. Based on the experimentally measured efficiency $\eta$, we can use Eq.~(\ref{eq:efficiency}) to estimate $g_{\text{eo}}$, and the results are presented in Extended Data Table~\ref{table:parameter}.

\vspace{0.4cm}
\noindent \textbf{Setup for noise measurement and photonic link}

\noindent The experimental setup for the added noise measurement and the 1-km photonic link is shown in Extended Data Fig.~\ref{Fig:twofridgesetup}. For added noise measurement, optical pulses are sent to Albert directly, bypassing Felix. The output channel 1, output channel 2, and input channel 1 of the LIA are disabled, and only the input channel 2 is enabled. 4.8~GHz microwave signals are used as the pump for TWPA, which provides $\sim 14$~dB gain. We first tune the VTS temperature and measure the output power spectral density in the absence of optical pulses to characterize the gain and added noise of the output chain. The LIA is used as a time-resolved electrical spectrum analyzer. The LIA has a low-pass filter bandwidth of 500~kHz, a sampling rate of 7~MHz, and a sampling time window of 15~\textmu s. At each frequency point, the LIA measures the averaged magnitude over 500 sequences, and the averaged power spectral density can be obtained by sweeping the local oscillator frequency. We then send optical pulses to Albert and measure the corresponding power spectral density at different optical peak power levels to characterize the light-induced time-resolved added noise spectrum. 

For the 1-km photonic link experiment, the optical pulses are sent to Felix, and the output pulses from Felix are sent to a 1-km fiber spool and then to Albert. Two DC voltages are applied to each transducer, and the voltages are fine tuned to achieve intra-cavity and inter-cavity frequency matching. The laser frequency is tuned to the pump frequency $\omega_{-,\text{Felix}}$, and an optical SSBM driven at $ \omega_{-,\text{Albert}} - \omega_{-,\text{Felix}}=2\pi \cdot 5.354$~GHz is used to generate an optical sideband as the pump for Albert at $\omega_{-,\text{Albert}}$. The power of the sideband at $\omega_{-,\text{Albert}}$ is 10~dB lower than the power at $\omega_{-,\text{Felix}}$. Therefore, at Felix, the optical power is dominated by the pump for Felix. Because the optical red sideband resonance mode of Felix at $\omega_{-,\text{Felix}}$ has an extinction ratio $\sim 20$~dB, the optical power is dominated by the pump at $\omega_{-,\text{Albert}}$ for Albert. The microwave pump for TWPA is disabled for photonic link experiment. To measure the M2O2M scattering matrix, the signals from LIA output channel 1 is up-converted to $~\sim 9.960$~GHz and sent to Felix. The microwave output from Albert is down-converted by a mixer whose local oscillator frequency is $~\sim 4.606$~GHz and subsequently sent to LIA input channel 2. The 4.606~GHz, 9.960~GHz, and 5.354~GHz signal generators are mutually phase locked by connecting their 10~MHz reference frequency. The M2O2M scattering matrix spectrum is obtained by sweeping the LIA's frequency. When performing the quadrature measurement using QPSK, we tune the phase of output channel 1, and collect 50 individual pulses at the input channel 2 for each phase. When measuring the interference pattern, we enable the output channel 2 and adjust its amplitude to match that of the received signals at input channel 2. We sweep the phase of output channel 2 and record the corresponding microwave power at input channel 2.

\section*{{\normalsize{}Acknowledgments}}

{\noindent\footnotesize{}This work is funded by the Co-design Center for Quantum Advantage (DE-SC0012704). The authors would like to thank Dr. Xu Han, Dr. Sihao Wang, and Dr. Wei Fu for refrigeration hardware installation, and Dr. Yong Sun, Dr. Lauren McCabe, Kelly Woods, Dr. Yeongjae Shin, Dr. Michael Rooks, and Dr. Sungwoo Sohn for their assistance provided in the device fabrication. The fabrication of the devices was done at the Yale School of Engineering \& Applied Science (SEAS) Cleanroom and the Yale Institute for Nanoscience and Quantum Engineering (YINQE). The TWPA used in this experiment is provided by IARPA and MIT Lincoln Laboratory.}{\footnotesize\par}

\section*{{\normalsize{}Author contributions }}

{\noindent\footnotesize{}H.X.T. and Y.Z. conceived the idea and experiment; Y.Z. fabricated the devices with assistance from M.S., L.Y., and J.X.; Y.Z. performed the measurements with assistance from Y.W. and C.L.; Y.Z., Y.W., C.L., and H.X.T. analyzed the data; Y.Z. and H.X.T. wrote the manuscript with inputs from all authors; H.X.T. supervised the project.}{\footnotesize\par}

\section*{{\normalsize{}Competing interests }}

{\noindent\footnotesize{}The authors declare no competing interests. }{\footnotesize\par}

\section*{{\normalsize{}Additional information}}
{\noindent\footnotesize{}\textbf{Correspondence and requests for materials} should be addressed to H.X.T.}
{\footnotesize\par}


\setcounter{figure}{0}
\renewcommand{\figurename}{Extended Data Fig.}

\setcounter{table}{0}
\renewcommand{\tablename}{Extended Data Table}

\begin{figure*}[t]
\capstart
\centering
\includegraphics[width=\linewidth]{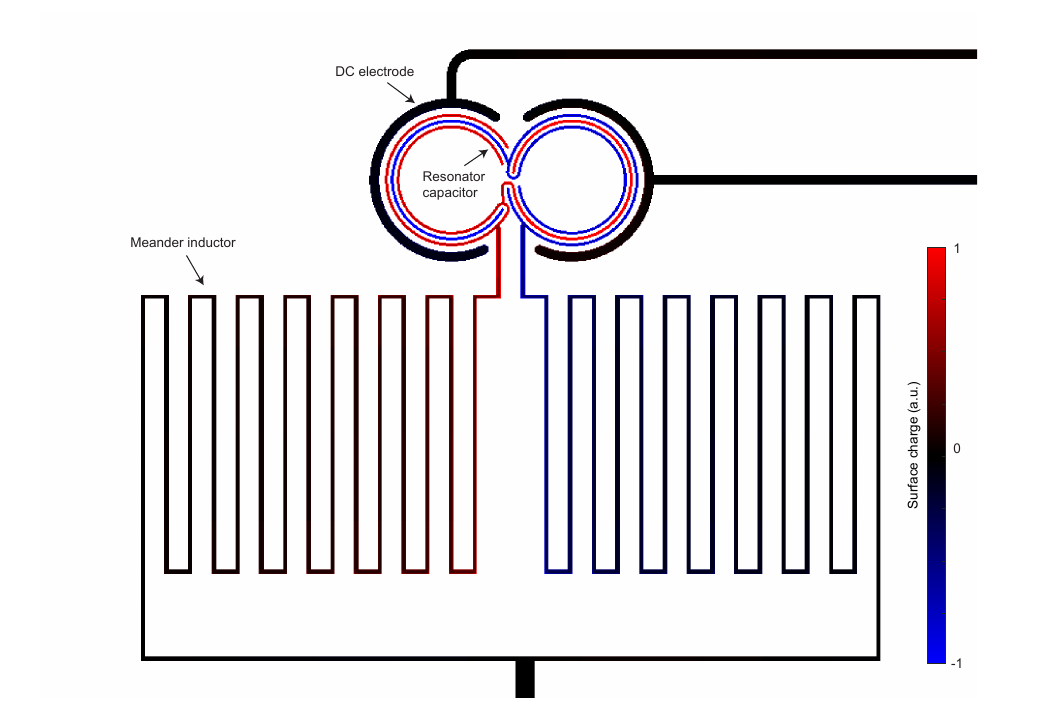}
\caption{\textbf{Surface charge distribution of the superconducting resonator.} The surface charge distribution of the superconducting resonator when excited with on-resonance microwave signals. The simulation is performed with Sonnet software. The charges are concentrated in the capacitor region, suggesting that the electric field is also confined at the capacitor. A $\pi$ phase can be seen between the two rings due to the reversed polarity.}
\label{Fig:charge}
\end{figure*}

\begin{figure*}[t]
\capstart
\centering
\includegraphics[width=\linewidth]{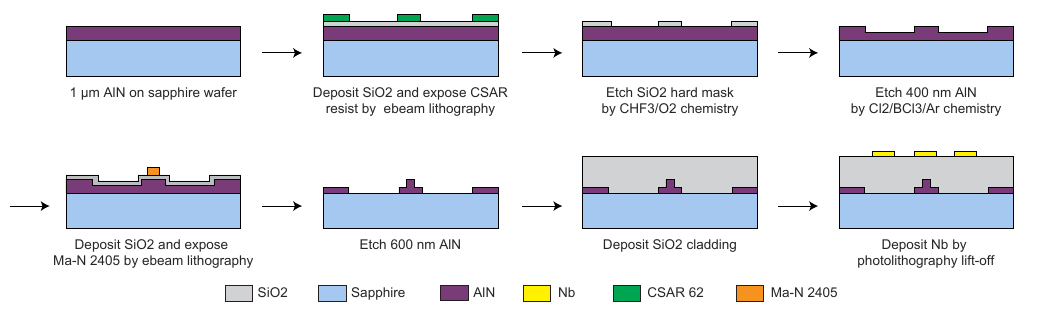}
\caption{\textbf{Fabrication process flow of an AlN transducer.} See Methods for detailed descriptions.}
\label{Fig:fabflow}
\end{figure*}

\begin{figure*}[t]
\capstart
\centering
\includegraphics[width=\linewidth]{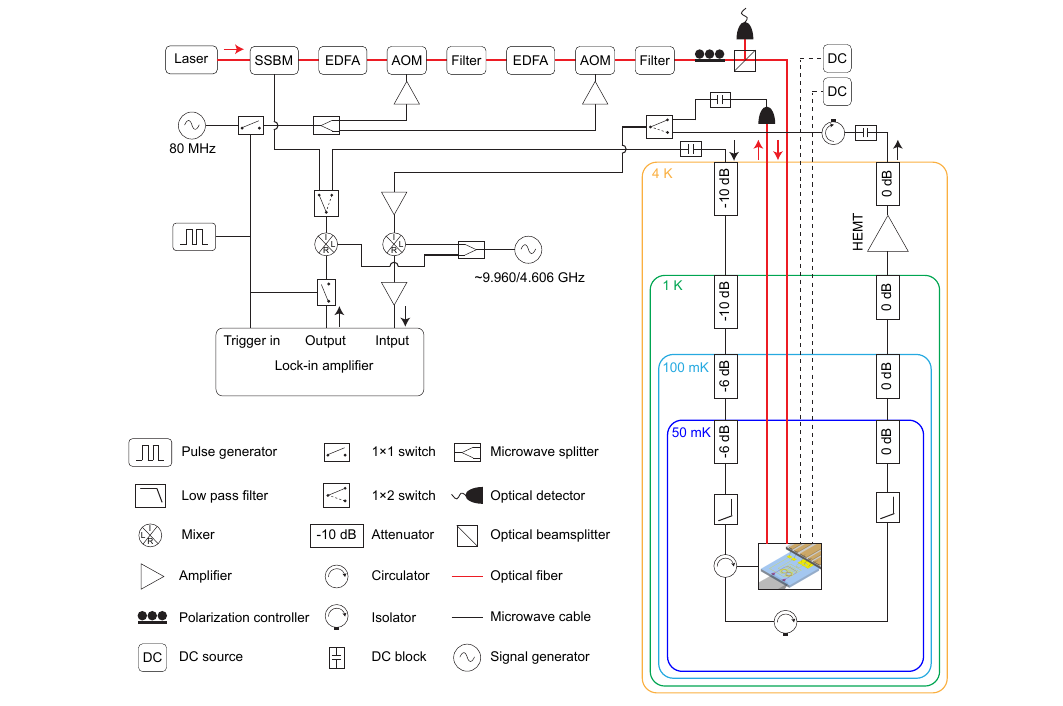}
\caption{\textbf{Experimental setup for transduction efficiency characterization.} See Methods for detailed descriptions.}
\label{Fig:conversion_setup}
\end{figure*}

\begin{figure*}[t]
\capstart
\centering
\includegraphics[width=\linewidth]{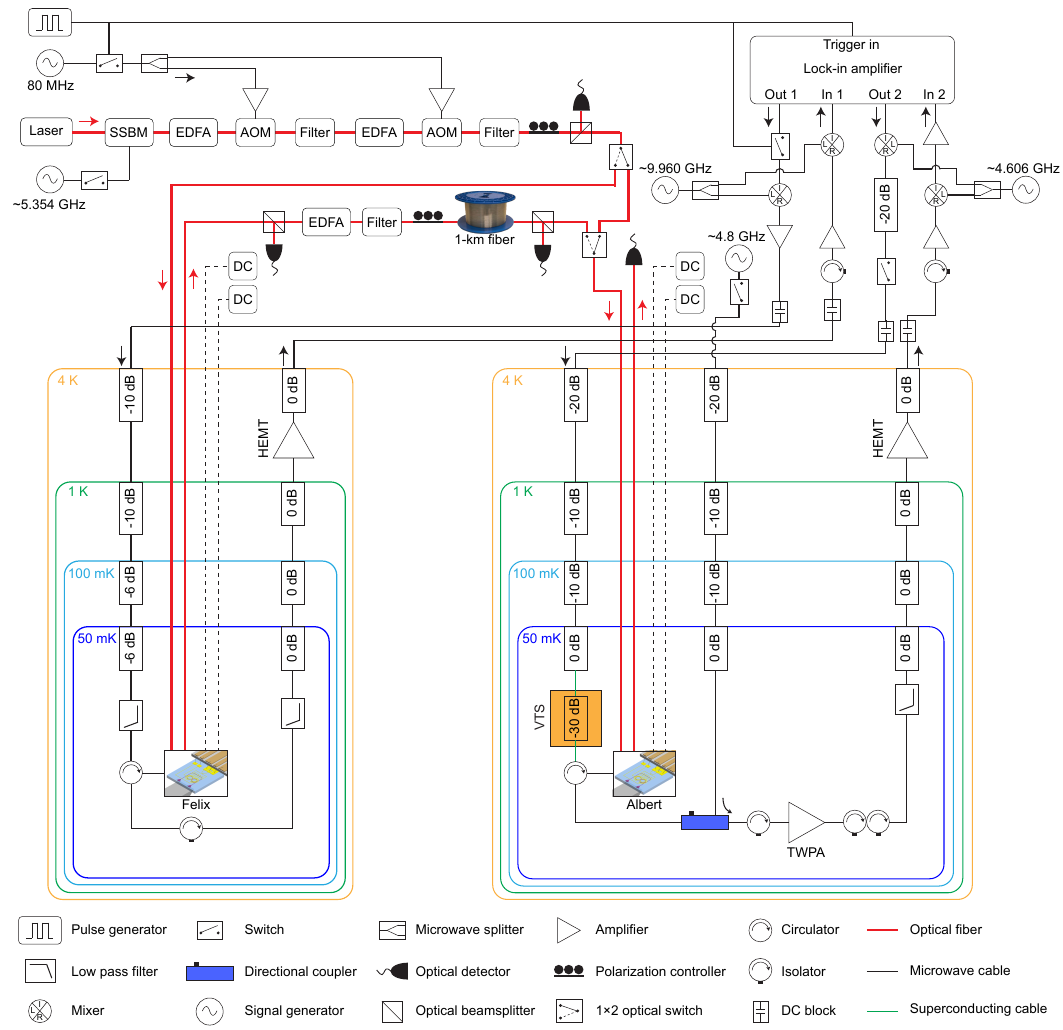}
\caption{\textbf{Experimental setup for noise measurement and 1-km photonic link.} See Methods for detailed descriptions.}
\label{Fig:twofridgesetup}
\end{figure*}

\clearpage

\setlength{\tabcolsep}{10pt} 
\renewcommand{\arraystretch}{1.5} 

\begin{table*}[]
\begin{center}
\begin{tabular}{|l|c|c|}
\hline
\multicolumn{1}{|c|}{Parameter}                                                                     & Felix                     & Albert                    \\ \hline
Optical red sideband mode frequency $\omega_{-}$                                                    & $2\pi \cdot 190.6320$ THz & $2\pi \cdot 190.6374$ THz \\ \hline
Optical red sideband mode intrinsic loss rate $\kappa_{-, \text{in}}$                               & $2\pi \cdot 134$ MHz      & $2\pi \cdot 214$ MHz      \\ \hline
Optical red sideband mode external loss rate $\kappa_{-, \text{ex}}$                                & $2\pi \cdot 102$ MHz      & $2\pi \cdot 77$ MHz       \\ \hline
Optical red sideband mode total loss rate $\kappa_{-}=\kappa_{-,\text{in}} + \kappa_{-,\text{ex}}$  & $2\pi \cdot 236$ MHz      & $2\pi \cdot 291$ MHz      \\ \hline
Optical blue sideband mode frequency $\omega_{+}$                                                   & $2\pi \cdot 190.6420$ THz & $2\pi \cdot 190.6420$ THz \\ \hline
Optical blue sideband mode intrinsic loss rate $\kappa_{+, \text{in}}$                              & $2\pi \cdot 118$ MHz      & $2\pi \cdot 167$ MHz      \\ \hline
Optical blue sideband mode external loss rate $\kappa_{+, \text{ex}}$                               & $2\pi \cdot 90$ MHz       & $2\pi \cdot 50$ MHz       \\ \hline
Optical blue sideband mode total loss rate $\kappa_{+}=\kappa_{+,\text{in}} + \kappa_{+,\text{ex}}$ & $2\pi \cdot 208$ MHz      & $2\pi \cdot 217$ MHz      \\ \hline
Microwave mode frequency $\omega_{\text{m}}$                                                                 & $2\pi \cdot 9.960$ GHz    & $2\pi \cdot 4.606$ GHz    \\ \hline
Microwave mode intrinsic loss rate $\kappa_{\text{m,in}}$                                           & $2\pi \cdot 23.3$ MHz     & $2\pi \cdot 2.4$ MHz      \\ \hline
Microwave mode external loss rate $\kappa_{\text{m,ex}}$                                            & $2\pi \cdot 14.7$ MHz     & $2\pi \cdot 11.5$ MHz     \\ \hline
Microwave mode total loss rate $\kappa_{\text{m}}=\kappa_{\text{m,in}} + \kappa_{\text{m,ex}} $     & $2\pi \cdot 38.0$ MHz     & $2\pi \cdot 13.9$ MHz     \\ \hline
Experimentally estimated $g_{\text{eo}}$                  & $2\pi \cdot 283$ Hz       & $2\pi \cdot 275$ Hz       \\ \hline
\end{tabular}
\caption{\textbf{Experimentally measured parameters for electro-optic transducers.} See Methods for the estimation of the electro-optic single-photon coupling rate $g_{\text{eo}}$.}
\label{table:parameter}
\end{center}
\end{table*}


\begin{thebibliography}{54}%
\makeatletter
\providecommand \@ifxundefined [1]{%
 \@ifx{#1\undefined}
}%
\providecommand \@ifnum [1]{%
 \ifnum #1\expandafter \@firstoftwo
 \else \expandafter \@secondoftwo
 \fi
}%
\providecommand \@ifx [1]{%
 \ifx #1\expandafter \@firstoftwo
 \else \expandafter \@secondoftwo
 \fi
}%
\providecommand \natexlab [1]{#1}%
\providecommand \enquote  [1]{``#1''}%
\providecommand \bibnamefont  [1]{#1}%
\providecommand \bibfnamefont [1]{#1}%
\providecommand \citenamefont [1]{#1}%
\providecommand \href@noop [0]{\@secondoftwo}%
\providecommand \href [0]{\begingroup \@sanitize@url \@href}%
\providecommand \@href[1]{\@@startlink{#1}\@@href}%
\providecommand \@@href[1]{\endgroup#1\@@endlink}%
\providecommand \@sanitize@url [0]{\catcode `\\12\catcode `\$12\catcode `\&12\catcode `\#12\catcode `\^12\catcode `\_12\catcode `\%12\relax}%
\providecommand \@@startlink[1]{}%
\providecommand \@@endlink[0]{}%
\providecommand \url  [0]{\begingroup\@sanitize@url \@url }%
\providecommand \@url [1]{\endgroup\@href {#1}{\urlprefix }}%
\providecommand \urlprefix  [0]{URL }%
\providecommand \Eprint [0]{\href }%
\providecommand \doibase [0]{https://doi.org/}%
\providecommand \selectlanguage [0]{\@gobble}%
\providecommand \bibinfo  [0]{\@secondoftwo}%
\providecommand \bibfield  [0]{\@secondoftwo}%
\providecommand \translation [1]{[#1]}%
\providecommand \BibitemOpen [0]{}%
\providecommand \bibitemStop [0]{}%
\providecommand \bibitemNoStop [0]{.\EOS\space}%
\providecommand \EOS [0]{\spacefactor3000\relax}%
\providecommand \BibitemShut  [1]{\csname bibitem#1\endcsname}%
\let\auto@bib@innerbib\@empty
\bibitem [{\citenamefont {Neill}\ \emph {et~al.}(2018)\citenamefont {Neill}, \citenamefont {Roushan}, \citenamefont {Kechedzhi}, \citenamefont {Boixo}, \citenamefont {Isakov}, \citenamefont {Smelyanskiy}, \citenamefont {Megrant}, \citenamefont {Chiaro}, \citenamefont {Dunsworth}, \citenamefont {Arya} \emph {et~al.}}]{neill2018blueprint}%
  \BibitemOpen
  \bibfield  {author} {\bibinfo {author} {\bibfnamefont {C.}~\bibnamefont {Neill}}, \bibinfo {author} {\bibfnamefont {P.}~\bibnamefont {Roushan}}, \bibinfo {author} {\bibfnamefont {K.}~\bibnamefont {Kechedzhi}}, \bibinfo {author} {\bibfnamefont {S.}~\bibnamefont {Boixo}}, \bibinfo {author} {\bibfnamefont {S.~V.}\ \bibnamefont {Isakov}}, \bibinfo {author} {\bibfnamefont {V.}~\bibnamefont {Smelyanskiy}}, \bibinfo {author} {\bibfnamefont {A.}~\bibnamefont {Megrant}}, \bibinfo {author} {\bibfnamefont {B.}~\bibnamefont {Chiaro}}, \bibinfo {author} {\bibfnamefont {A.}~\bibnamefont {Dunsworth}}, \bibinfo {author} {\bibfnamefont {K.}~\bibnamefont {Arya}}, \emph {et~al.},\ }\enquote{\bibinfo {title} {A blueprint for demonstrating quantum supremacy with superconducting qubits},} \href@noop {} {\bibfield  {journal} {\bibinfo  {journal} {\emph {Science}}}\ }\textbf {\bibinfo {volume} {360}},\ \bibinfo {pages} {195} (\bibinfo {year} {2018})\BibitemShut {NoStop}%
\bibitem [{\citenamefont {Acharya}\ \emph {et~al.}(2025)\citenamefont {Acharya}, \citenamefont {Abanin}, \citenamefont {Aghababaie-Beni}, \citenamefont {Aleiner}, \citenamefont {Andersen}, \citenamefont {Ansmann}, \citenamefont {Arute}, \citenamefont {Arya}, \citenamefont {Asfaw}, \citenamefont {Astrakhantsev} \emph {et~al.}}]{acharya2025quantum}%
  \BibitemOpen
  \bibfield  {author} {\bibinfo {author} {\bibfnamefont {R.}~\bibnamefont {Acharya}}, \bibinfo {author} {\bibfnamefont {D.~A.}\ \bibnamefont {Abanin}}, \bibinfo {author} {\bibfnamefont {L.}~\bibnamefont {Aghababaie-Beni}}, \bibinfo {author} {\bibfnamefont {I.}~\bibnamefont {Aleiner}}, \bibinfo {author} {\bibfnamefont {T.~I.}\ \bibnamefont {Andersen}}, \bibinfo {author} {\bibfnamefont {M.}~\bibnamefont {Ansmann}}, \bibinfo {author} {\bibfnamefont {F.}~\bibnamefont {Arute}}, \bibinfo {author} {\bibfnamefont {K.}~\bibnamefont {Arya}}, \bibinfo {author} {\bibfnamefont {A.}~\bibnamefont {Asfaw}}, \bibinfo {author} {\bibfnamefont {N.}~\bibnamefont {Astrakhantsev}}, \emph {et~al.},\ }\enquote{\bibinfo {title} {Quantum error correction below the surface code threshold},} \href@noop {} {\bibfield  {journal} {\bibinfo  {journal} {\emph {Nature}}}\ }\textbf {\bibinfo {volume} {638}},\ \bibinfo {pages} {920} (\bibinfo {year} {2025})\BibitemShut {NoStop}%
\bibitem [{\citenamefont {Gao}\ \emph {et~al.}(2025)\citenamefont {Gao}, \citenamefont {Fan}, \citenamefont {Zha}, \citenamefont {Bei}, \citenamefont {Cai}, \citenamefont {Cai}, \citenamefont {Cao}, \citenamefont {Chen}, \citenamefont {Chen}, \citenamefont {Chen} \emph {et~al.}}]{gao2025establishing}%
  \BibitemOpen
  \bibfield  {author} {\bibinfo {author} {\bibfnamefont {D.}~\bibnamefont {Gao}}, \bibinfo {author} {\bibfnamefont {D.}~\bibnamefont {Fan}}, \bibinfo {author} {\bibfnamefont {C.}~\bibnamefont {Zha}}, \bibinfo {author} {\bibfnamefont {J.}~\bibnamefont {Bei}}, \bibinfo {author} {\bibfnamefont {G.}~\bibnamefont {Cai}}, \bibinfo {author} {\bibfnamefont {J.}~\bibnamefont {Cai}}, \bibinfo {author} {\bibfnamefont {S.}~\bibnamefont {Cao}}, \bibinfo {author} {\bibfnamefont {F.}~\bibnamefont {Chen}}, \bibinfo {author} {\bibfnamefont {J.}~\bibnamefont {Chen}}, \bibinfo {author} {\bibfnamefont {K.}~\bibnamefont {Chen}}, \emph {et~al.},\ }\enquote{\bibinfo {title} {Establishing a new benchmark in quantum computational advantage with 105-qubit zuchongzhi 3.0 processor},} \href@noop {} {\bibfield  {journal} {\bibinfo  {journal} {\emph {Phys. Rev. Lett.}}}\ }\textbf {\bibinfo {volume} {134}},\ \bibinfo {pages} {090601} (\bibinfo {year} {2025})\BibitemShut {NoStop}%
\bibitem [{\citenamefont {Boixo}\ \emph {et~al.}(2014)\citenamefont {Boixo}, \citenamefont {R{\o}nnow}, \citenamefont {Isakov}, \citenamefont {Wang}, \citenamefont {Wecker}, \citenamefont {Lidar}, \citenamefont {Martinis},\ and\ \citenamefont {Troyer}}]{boixo2014evidence}%
  \BibitemOpen
  \bibfield  {author} {\bibinfo {author} {\bibfnamefont {S.}~\bibnamefont {Boixo}}, \bibinfo {author} {\bibfnamefont {T.~F.}\ \bibnamefont {R{\o}nnow}}, \bibinfo {author} {\bibfnamefont {S.~V.}\ \bibnamefont {Isakov}}, \bibinfo {author} {\bibfnamefont {Z.}~\bibnamefont {Wang}}, \bibinfo {author} {\bibfnamefont {D.}~\bibnamefont {Wecker}}, \bibinfo {author} {\bibfnamefont {D.~A.}\ \bibnamefont {Lidar}}, \bibinfo {author} {\bibfnamefont {J.~M.}\ \bibnamefont {Martinis}},\ and\ \bibinfo {author} {\bibfnamefont {M.}~\bibnamefont {Troyer}},\ }\enquote{\bibinfo {title} {Evidence for quantum annealing with more than one hundred qubits},} \href@noop {} {\bibfield  {journal} {\bibinfo  {journal} {\emph {Nat. Phys.}}}\ }\textbf {\bibinfo {volume} {10}},\ \bibinfo {pages} {218} (\bibinfo {year} {2014})\BibitemShut {NoStop}%
\bibitem [{\citenamefont {Webber}\ \emph {et~al.}(2022)\citenamefont {Webber}, \citenamefont {Elfving}, \citenamefont {Weidt},\ and\ \citenamefont {Hensinger}}]{webber2022impact}%
  \BibitemOpen
  \bibfield  {author} {\bibinfo {author} {\bibfnamefont {M.}~\bibnamefont {Webber}}, \bibinfo {author} {\bibfnamefont {V.}~\bibnamefont {Elfving}}, \bibinfo {author} {\bibfnamefont {S.}~\bibnamefont {Weidt}},\ and\ \bibinfo {author} {\bibfnamefont {W.~K.}\ \bibnamefont {Hensinger}},\ }\enquote{\bibinfo {title} {The impact of hardware specifications on reaching quantum advantage in the fault tolerant regime},} \href@noop {} {\bibfield  {journal} {\bibinfo  {journal} {\emph {AVS Quantum Sci.}}}\ }\textbf {\bibinfo {volume} {4}} (\bibinfo {year} {2022})\BibitemShut {NoStop}%
\bibitem [{\citenamefont {Krinner}\ \emph {et~al.}(2019)\citenamefont {Krinner}, \citenamefont {Storz}, \citenamefont {Kurpiers}, \citenamefont {Magnard}, \citenamefont {Heinsoo}, \citenamefont {Keller}, \citenamefont {Luetolf}, \citenamefont {Eichler},\ and\ \citenamefont {Wallraff}}]{krinner2019engineering}%
  \BibitemOpen
  \bibfield  {author} {\bibinfo {author} {\bibfnamefont {S.}~\bibnamefont {Krinner}}, \bibinfo {author} {\bibfnamefont {S.}~\bibnamefont {Storz}}, \bibinfo {author} {\bibfnamefont {P.}~\bibnamefont {Kurpiers}}, \bibinfo {author} {\bibfnamefont {P.}~\bibnamefont {Magnard}}, \bibinfo {author} {\bibfnamefont {J.}~\bibnamefont {Heinsoo}}, \bibinfo {author} {\bibfnamefont {R.}~\bibnamefont {Keller}}, \bibinfo {author} {\bibfnamefont {J.}~\bibnamefont {Luetolf}}, \bibinfo {author} {\bibfnamefont {C.}~\bibnamefont {Eichler}},\ and\ \bibinfo {author} {\bibfnamefont {A.}~\bibnamefont {Wallraff}},\ }\enquote{\bibinfo {title} {Engineering cryogenic setups for 100-qubit scale superconducting circuit systems},} \href@noop {} {\bibfield  {journal} {\bibinfo  {journal} {\emph {EPJ Quantum Technol.}}}\ }\textbf {\bibinfo {volume} {6}},\ \bibinfo {pages} {2} (\bibinfo {year} {2019})\BibitemShut {NoStop}%
\bibitem [{\citenamefont {Magnard}\ \emph {et~al.}(2020)\citenamefont {Magnard}, \citenamefont {Storz}, \citenamefont {Kurpiers}, \citenamefont {Sch{\"a}r}, \citenamefont {Marxer}, \citenamefont {L{\"u}tolf}, \citenamefont {Walter}, \citenamefont {Besse}, \citenamefont {Gabureac}, \citenamefont {Reuer} \emph {et~al.}}]{magnard2020microwave}%
  \BibitemOpen
  \bibfield  {author} {\bibinfo {author} {\bibfnamefont {P.}~\bibnamefont {Magnard}}, \bibinfo {author} {\bibfnamefont {S.}~\bibnamefont {Storz}}, \bibinfo {author} {\bibfnamefont {P.}~\bibnamefont {Kurpiers}}, \bibinfo {author} {\bibfnamefont {J.}~\bibnamefont {Sch{\"a}r}}, \bibinfo {author} {\bibfnamefont {F.}~\bibnamefont {Marxer}}, \bibinfo {author} {\bibfnamefont {J.}~\bibnamefont {L{\"u}tolf}}, \bibinfo {author} {\bibfnamefont {T.}~\bibnamefont {Walter}}, \bibinfo {author} {\bibfnamefont {J.-C.}\ \bibnamefont {Besse}}, \bibinfo {author} {\bibfnamefont {M.}~\bibnamefont {Gabureac}}, \bibinfo {author} {\bibfnamefont {K.}~\bibnamefont {Reuer}}, \emph {et~al.},\ }\enquote{\bibinfo {title} {Microwave quantum link between superconducting circuits housed in spatially separated cryogenic systems},} \href@noop {} {\bibfield  {journal} {\bibinfo  {journal} {\emph {Phys. Rev. Lett.}}}\ }\textbf {\bibinfo {volume} {125}},\ \bibinfo {pages} {260502} (\bibinfo {year} {2020})\BibitemShut {NoStop}%
\bibitem [{\citenamefont {Yam}\ \emph {et~al.}(2025)\citenamefont {Yam}, \citenamefont {Renger}, \citenamefont {Gandorfer}, \citenamefont {Fesquet}, \citenamefont {Handschuh}, \citenamefont {Honasoge}, \citenamefont {Kronowetter}, \citenamefont {Nojiri}, \citenamefont {Partanen}, \citenamefont {Pfeiffer} \emph {et~al.}}]{yam2025cryogenic}%
  \BibitemOpen
  \bibfield  {author} {\bibinfo {author} {\bibfnamefont {W.}~\bibnamefont {Yam}}, \bibinfo {author} {\bibfnamefont {M.}~\bibnamefont {Renger}}, \bibinfo {author} {\bibfnamefont {S.}~\bibnamefont {Gandorfer}}, \bibinfo {author} {\bibfnamefont {F.}~\bibnamefont {Fesquet}}, \bibinfo {author} {\bibfnamefont {M.}~\bibnamefont {Handschuh}}, \bibinfo {author} {\bibfnamefont {K.}~\bibnamefont {Honasoge}}, \bibinfo {author} {\bibfnamefont {F.}~\bibnamefont {Kronowetter}}, \bibinfo {author} {\bibfnamefont {Y.}~\bibnamefont {Nojiri}}, \bibinfo {author} {\bibfnamefont {M.}~\bibnamefont {Partanen}}, \bibinfo {author} {\bibfnamefont {M.}~\bibnamefont {Pfeiffer}}, \emph {et~al.},\ }\enquote{\bibinfo {title} {Cryogenic microwave link for quantum local area networks},} \href@noop {} {\bibfield  {journal} {\bibinfo  {journal} {\emph {npj Quantum Info.}}}\ }\textbf {\bibinfo {volume} {11}},\ \bibinfo {pages} {1} (\bibinfo {year} {2025})\BibitemShut {NoStop}%
\bibitem [{\citenamefont {Wooten}\ \emph {et~al.}(2000)\citenamefont {Wooten}, \citenamefont {Kissa}, \citenamefont {Yi-Yan}, \citenamefont {Murphy}, \citenamefont {Lafaw}, \citenamefont {Hallemeier}, \citenamefont {Maack}, \citenamefont {Attanasio}, \citenamefont {Fritz}, \citenamefont {McBrien} \emph {et~al.}}]{wooten2000review}%
  \BibitemOpen
  \bibfield  {author} {\bibinfo {author} {\bibfnamefont {E.~L.}\ \bibnamefont {Wooten}}, \bibinfo {author} {\bibfnamefont {K.~M.}\ \bibnamefont {Kissa}}, \bibinfo {author} {\bibfnamefont {A.}~\bibnamefont {Yi-Yan}}, \bibinfo {author} {\bibfnamefont {E.~J.}\ \bibnamefont {Murphy}}, \bibinfo {author} {\bibfnamefont {D.~A.}\ \bibnamefont {Lafaw}}, \bibinfo {author} {\bibfnamefont {P.~F.}\ \bibnamefont {Hallemeier}}, \bibinfo {author} {\bibfnamefont {D.}~\bibnamefont {Maack}}, \bibinfo {author} {\bibfnamefont {D.~V.}\ \bibnamefont {Attanasio}}, \bibinfo {author} {\bibfnamefont {D.~J.}\ \bibnamefont {Fritz}}, \bibinfo {author} {\bibfnamefont {G.~J.}\ \bibnamefont {McBrien}}, \emph {et~al.},\ }\enquote{\bibinfo {title} {A review of lithium niobate modulators for fiber-optic communications systems},} \href@noop {} {\bibfield  {journal} {\bibinfo  {journal} {\emph {IEEE J. Sel. Top. Quantum Electron.}}}\ }\textbf {\bibinfo {volume} {6}},\ \bibinfo {pages} {69} (\bibinfo {year} {2000})\BibitemShut {NoStop}%
\bibitem [{\citenamefont {Youssefi}\ \emph {et~al.}(2021)\citenamefont {Youssefi}, \citenamefont {Shomroni}, \citenamefont {Joshi}, \citenamefont {Bernier}, \citenamefont {Lukashchuk}, \citenamefont {Uhrich}, \citenamefont {Qiu},\ and\ \citenamefont {Kippenberg}}]{youssefi2021cryogenic}%
  \BibitemOpen
  \bibfield  {author} {\bibinfo {author} {\bibfnamefont {A.}~\bibnamefont {Youssefi}}, \bibinfo {author} {\bibfnamefont {I.}~\bibnamefont {Shomroni}}, \bibinfo {author} {\bibfnamefont {Y.~J.}\ \bibnamefont {Joshi}}, \bibinfo {author} {\bibfnamefont {N.~R.}\ \bibnamefont {Bernier}}, \bibinfo {author} {\bibfnamefont {A.}~\bibnamefont {Lukashchuk}}, \bibinfo {author} {\bibfnamefont {P.}~\bibnamefont {Uhrich}}, \bibinfo {author} {\bibfnamefont {L.}~\bibnamefont {Qiu}},\ and\ \bibinfo {author} {\bibfnamefont {T.~J.}\ \bibnamefont {Kippenberg}},\ }\enquote{\bibinfo {title} {A cryogenic electro-optic interconnect for superconducting devices},} \href@noop {} {\bibfield  {journal} {\bibinfo  {journal} {\emph {Nat. Electron.}}}\ }\textbf {\bibinfo {volume} {4}},\ \bibinfo {pages} {326} (\bibinfo {year} {2021})\BibitemShut {NoStop}%
\bibitem [{\citenamefont {Shen}\ \emph {et~al.}(2024)\citenamefont {Shen}, \citenamefont {Xie}, \citenamefont {Xu}, \citenamefont {Wang}, \citenamefont {Cheng}, \citenamefont {Fu}, \citenamefont {Zhou},\ and\ \citenamefont {Tang}}]{shen2024photonic}%
  \BibitemOpen
  \bibfield  {author} {\bibinfo {author} {\bibfnamefont {M.}~\bibnamefont {Shen}}, \bibinfo {author} {\bibfnamefont {J.}~\bibnamefont {Xie}}, \bibinfo {author} {\bibfnamefont {Y.}~\bibnamefont {Xu}}, \bibinfo {author} {\bibfnamefont {S.}~\bibnamefont {Wang}}, \bibinfo {author} {\bibfnamefont {R.}~\bibnamefont {Cheng}}, \bibinfo {author} {\bibfnamefont {W.}~\bibnamefont {Fu}}, \bibinfo {author} {\bibfnamefont {Y.}~\bibnamefont {Zhou}},\ and\ \bibinfo {author} {\bibfnamefont {H.~X.}\ \bibnamefont {Tang}},\ }\enquote{\bibinfo {title} {Photonic link from single-flux-quantum circuits to room temperature},} \href@noop {} {\bibfield  {journal} {\bibinfo  {journal} {\emph {Nat. Photon.}}}\ }\textbf {\bibinfo {volume} {18}},\ \bibinfo {pages} {371} (\bibinfo {year} {2024})\BibitemShut {NoStop}%
\bibitem [{\citenamefont {Fan}\ \emph {et~al.}(2018)\citenamefont {Fan}, \citenamefont {Zou}, \citenamefont {Cheng}, \citenamefont {Guo}, \citenamefont {Han}, \citenamefont {Gong}, \citenamefont {Wang},\ and\ \citenamefont {Tang}}]{fan2018superconducting}%
  \BibitemOpen
  \bibfield  {author} {\bibinfo {author} {\bibfnamefont {L.}~\bibnamefont {Fan}}, \bibinfo {author} {\bibfnamefont {C.-L.}\ \bibnamefont {Zou}}, \bibinfo {author} {\bibfnamefont {R.}~\bibnamefont {Cheng}}, \bibinfo {author} {\bibfnamefont {X.}~\bibnamefont {Guo}}, \bibinfo {author} {\bibfnamefont {X.}~\bibnamefont {Han}}, \bibinfo {author} {\bibfnamefont {Z.}~\bibnamefont {Gong}}, \bibinfo {author} {\bibfnamefont {S.}~\bibnamefont {Wang}},\ and\ \bibinfo {author} {\bibfnamefont {H.~X.}\ \bibnamefont {Tang}},\ }\enquote{\bibinfo {title} {Superconducting cavity electro-optics: a platform for coherent photon conversion between superconducting and photonic circuits},} \href@noop {} {\bibfield  {journal} {\bibinfo  {journal} {\emph {Sci. Adv.}}}\ }\textbf {\bibinfo {volume} {4}},\ \bibinfo {pages} {eaar4994} (\bibinfo {year} {2018})\BibitemShut {NoStop}%
\bibitem [{\citenamefont {Arnold}\ \emph {et~al.}(2025)\citenamefont {Arnold}, \citenamefont {Werner}, \citenamefont {Sahu}, \citenamefont {Kapoor}, \citenamefont {Qiu},\ and\ \citenamefont {Fink}}]{arnold2025all}%
  \BibitemOpen
  \bibfield  {author} {\bibinfo {author} {\bibfnamefont {G.}~\bibnamefont {Arnold}}, \bibinfo {author} {\bibfnamefont {T.}~\bibnamefont {Werner}}, \bibinfo {author} {\bibfnamefont {R.}~\bibnamefont {Sahu}}, \bibinfo {author} {\bibfnamefont {L.~N.}\ \bibnamefont {Kapoor}}, \bibinfo {author} {\bibfnamefont {L.}~\bibnamefont {Qiu}},\ and\ \bibinfo {author} {\bibfnamefont {J.~M.}\ \bibnamefont {Fink}},\ }\enquote{\bibinfo {title} {All-optical superconducting qubit readout},} \href@noop {} {\bibfield  {journal} {\bibinfo  {journal} {\emph {Nat. Phys.}}}\ }\textbf {\bibinfo {volume} {21}},\ \bibinfo {pages} {393} (\bibinfo {year} {2025})\BibitemShut {NoStop}%
\bibitem [{\citenamefont {Warner}\ \emph {et~al.}(2025)\citenamefont {Warner}, \citenamefont {Holzgrafe}, \citenamefont {Yankelevich}, \citenamefont {Barton}, \citenamefont {Poletto}, \citenamefont {Xin}, \citenamefont {Sinclair}, \citenamefont {Zhu}, \citenamefont {Sete}, \citenamefont {Langley} \emph {et~al.}}]{warner2025coherent}%
  \BibitemOpen
  \bibfield  {author} {\bibinfo {author} {\bibfnamefont {H.~K.}\ \bibnamefont {Warner}}, \bibinfo {author} {\bibfnamefont {J.}~\bibnamefont {Holzgrafe}}, \bibinfo {author} {\bibfnamefont {B.}~\bibnamefont {Yankelevich}}, \bibinfo {author} {\bibfnamefont {D.}~\bibnamefont {Barton}}, \bibinfo {author} {\bibfnamefont {S.}~\bibnamefont {Poletto}}, \bibinfo {author} {\bibfnamefont {C.}~\bibnamefont {Xin}}, \bibinfo {author} {\bibfnamefont {N.}~\bibnamefont {Sinclair}}, \bibinfo {author} {\bibfnamefont {D.}~\bibnamefont {Zhu}}, \bibinfo {author} {\bibfnamefont {E.}~\bibnamefont {Sete}}, \bibinfo {author} {\bibfnamefont {B.}~\bibnamefont {Langley}}, \emph {et~al.},\ }\enquote{\bibinfo {title} {Coherent control of a superconducting qubit using light},} \href@noop {} {\bibfield  {journal} {\bibinfo  {journal} {\emph {Nat. Phys.}}}\ }\textbf {\bibinfo {volume} {21}},\ \bibinfo {pages} {831} (\bibinfo {year} {2025})\BibitemShut {NoStop}%
\bibitem [{\citenamefont {Xu}\ \emph {et~al.}(2021)\citenamefont {Xu}, \citenamefont {Sayem}, \citenamefont {Fan}, \citenamefont {Zou}, \citenamefont {Wang}, \citenamefont {Cheng}, \citenamefont {Fu}, \citenamefont {Yang}, \citenamefont {Xu},\ and\ \citenamefont {Tang}}]{xu2021bidirectional}%
  \BibitemOpen
  \bibfield  {author} {\bibinfo {author} {\bibfnamefont {Y.}~\bibnamefont {Xu}}, \bibinfo {author} {\bibfnamefont {A.~A.}\ \bibnamefont {Sayem}}, \bibinfo {author} {\bibfnamefont {L.}~\bibnamefont {Fan}}, \bibinfo {author} {\bibfnamefont {C.-L.}\ \bibnamefont {Zou}}, \bibinfo {author} {\bibfnamefont {S.}~\bibnamefont {Wang}}, \bibinfo {author} {\bibfnamefont {R.}~\bibnamefont {Cheng}}, \bibinfo {author} {\bibfnamefont {W.}~\bibnamefont {Fu}}, \bibinfo {author} {\bibfnamefont {L.}~\bibnamefont {Yang}}, \bibinfo {author} {\bibfnamefont {M.}~\bibnamefont {Xu}},\ and\ \bibinfo {author} {\bibfnamefont {H.~X.}\ \bibnamefont {Tang}},\ }\enquote{\bibinfo {title} {Bidirectional interconversion of microwave and light with thin-film lithium niobate},} \href@noop {} {\bibfield  {journal} {\bibinfo  {journal} {\emph {Nat. Commun.}}}\ }\textbf {\bibinfo {volume} {12}},\ \bibinfo {pages} {4453} (\bibinfo {year} {2021})\BibitemShut {NoStop}%
\bibitem [{\citenamefont {McKenna}\ \emph {et~al.}(2020)\citenamefont {McKenna}, \citenamefont {Witmer}, \citenamefont {Patel}, \citenamefont {Jiang}, \citenamefont {Van~Laer}, \citenamefont {Arrangoiz-Arriola}, \citenamefont {Wollack}, \citenamefont {Herrmann},\ and\ \citenamefont {Safavi-Naeini}}]{mckenna2020cryogenic}%
  \BibitemOpen
  \bibfield  {author} {\bibinfo {author} {\bibfnamefont {T.~P.}\ \bibnamefont {McKenna}}, \bibinfo {author} {\bibfnamefont {J.~D.}\ \bibnamefont {Witmer}}, \bibinfo {author} {\bibfnamefont {R.~N.}\ \bibnamefont {Patel}}, \bibinfo {author} {\bibfnamefont {W.}~\bibnamefont {Jiang}}, \bibinfo {author} {\bibfnamefont {R.}~\bibnamefont {Van~Laer}}, \bibinfo {author} {\bibfnamefont {P.}~\bibnamefont {Arrangoiz-Arriola}}, \bibinfo {author} {\bibfnamefont {E.~A.}\ \bibnamefont {Wollack}}, \bibinfo {author} {\bibfnamefont {J.~F.}\ \bibnamefont {Herrmann}},\ and\ \bibinfo {author} {\bibfnamefont {A.~H.}\ \bibnamefont {Safavi-Naeini}},\ }\enquote{\bibinfo {title} {Cryogenic microwave-to-optical conversion using a triply resonant lithium-niobate-on-sapphire transducer},} \href@noop {} {\bibfield  {journal} {\bibinfo  {journal} {\emph {Optica}}}\ }\textbf {\bibinfo {volume} {7}},\ \bibinfo {pages} {1737} (\bibinfo {year} {2020})\BibitemShut {NoStop}%
\bibitem [{\citenamefont {Pintus}\ \emph {et~al.}(2022{\natexlab{a}})\citenamefont {Pintus}, \citenamefont {Singh}, \citenamefont {Xie}, \citenamefont {Ranzani}, \citenamefont {Gustafsson}, \citenamefont {Tran}, \citenamefont {Xiang}, \citenamefont {Peters}, \citenamefont {Bowers},\ and\ \citenamefont {Soltani}}]{pintus2022ultralow}%
  \BibitemOpen
  \bibfield  {author} {\bibinfo {author} {\bibfnamefont {P.}~\bibnamefont {Pintus}}, \bibinfo {author} {\bibfnamefont {A.}~\bibnamefont {Singh}}, \bibinfo {author} {\bibfnamefont {W.}~\bibnamefont {Xie}}, \bibinfo {author} {\bibfnamefont {L.}~\bibnamefont {Ranzani}}, \bibinfo {author} {\bibfnamefont {M.~V.}\ \bibnamefont {Gustafsson}}, \bibinfo {author} {\bibfnamefont {M.~A.}\ \bibnamefont {Tran}}, \bibinfo {author} {\bibfnamefont {C.}~\bibnamefont {Xiang}}, \bibinfo {author} {\bibfnamefont {J.}~\bibnamefont {Peters}}, \bibinfo {author} {\bibfnamefont {J.~E.}\ \bibnamefont {Bowers}},\ and\ \bibinfo {author} {\bibfnamefont {M.}~\bibnamefont {Soltani}},\ }\enquote{\bibinfo {title} {Ultralow voltage, high-speed, and energy-efficient cryogenic electro-optic modulator},} \href@noop {} {\bibfield  {journal} {\bibinfo  {journal} {\emph {Optica}}}\ }\textbf {\bibinfo {volume} {9}},\ \bibinfo {pages} {1176} (\bibinfo {year} {2022}{\natexlab{a}})\BibitemShut {NoStop}%
\bibitem [{\citenamefont {Delaney}\ \emph {et~al.}(2022)\citenamefont {Delaney}, \citenamefont {Urmey}, \citenamefont {Mittal}, \citenamefont {Brubaker}, \citenamefont {Kindem}, \citenamefont {Burns}, \citenamefont {Regal},\ and\ \citenamefont {Lehnert}}]{delaney2022superconducting}%
  \BibitemOpen
  \bibfield  {author} {\bibinfo {author} {\bibfnamefont {R.}~\bibnamefont {Delaney}}, \bibinfo {author} {\bibfnamefont {M.}~\bibnamefont {Urmey}}, \bibinfo {author} {\bibfnamefont {S.}~\bibnamefont {Mittal}}, \bibinfo {author} {\bibfnamefont {B.}~\bibnamefont {Brubaker}}, \bibinfo {author} {\bibfnamefont {J.}~\bibnamefont {Kindem}}, \bibinfo {author} {\bibfnamefont {P.}~\bibnamefont {Burns}}, \bibinfo {author} {\bibfnamefont {C.}~\bibnamefont {Regal}},\ and\ \bibinfo {author} {\bibfnamefont {K.}~\bibnamefont {Lehnert}},\ }\enquote{\bibinfo {title} {Superconducting-qubit readout via low-backaction electro-optic transduction},} \href@noop {} {\bibfield  {journal} {\bibinfo  {journal} {\emph {Nature}}}\ }\textbf {\bibinfo {volume} {606}},\ \bibinfo {pages} {489} (\bibinfo {year} {2022})\BibitemShut {NoStop}%
\bibitem [{\citenamefont {Holzgrafe}\ \emph {et~al.}(2020)\citenamefont {Holzgrafe}, \citenamefont {Sinclair}, \citenamefont {Zhu}, \citenamefont {Shams-Ansari}, \citenamefont {Colangelo}, \citenamefont {Hu}, \citenamefont {Zhang}, \citenamefont {Berggren},\ and\ \citenamefont {Lon{\v{c}}ar}}]{holzgrafe2020cavity}%
  \BibitemOpen
  \bibfield  {author} {\bibinfo {author} {\bibfnamefont {J.}~\bibnamefont {Holzgrafe}}, \bibinfo {author} {\bibfnamefont {N.}~\bibnamefont {Sinclair}}, \bibinfo {author} {\bibfnamefont {D.}~\bibnamefont {Zhu}}, \bibinfo {author} {\bibfnamefont {A.}~\bibnamefont {Shams-Ansari}}, \bibinfo {author} {\bibfnamefont {M.}~\bibnamefont {Colangelo}}, \bibinfo {author} {\bibfnamefont {Y.}~\bibnamefont {Hu}}, \bibinfo {author} {\bibfnamefont {M.}~\bibnamefont {Zhang}}, \bibinfo {author} {\bibfnamefont {K.~K.}\ \bibnamefont {Berggren}},\ and\ \bibinfo {author} {\bibfnamefont {M.}~\bibnamefont {Lon{\v{c}}ar}},\ }\enquote{\bibinfo {title} {Cavity electro-optics in thin-film lithium niobate for efficient microwave-to-optical transduction},} \href@noop {} {\bibfield  {journal} {\bibinfo  {journal} {\emph {Optica}}}\ }\textbf {\bibinfo {volume} {7}},\ \bibinfo {pages} {1714} (\bibinfo {year} {2020})\BibitemShut {NoStop}%
\bibitem [{\citenamefont {Rueda}\ \emph {et~al.}(2016)\citenamefont {Rueda}, \citenamefont {Sedlmeir}, \citenamefont {Collodo}, \citenamefont {Vogl}, \citenamefont {Stiller}, \citenamefont {Schunk}, \citenamefont {Strekalov}, \citenamefont {Marquardt}, \citenamefont {Fink}, \citenamefont {Painter} \emph {et~al.}}]{rueda2016efficient}%
  \BibitemOpen
  \bibfield  {author} {\bibinfo {author} {\bibfnamefont {A.}~\bibnamefont {Rueda}}, \bibinfo {author} {\bibfnamefont {F.}~\bibnamefont {Sedlmeir}}, \bibinfo {author} {\bibfnamefont {M.~C.}\ \bibnamefont {Collodo}}, \bibinfo {author} {\bibfnamefont {U.}~\bibnamefont {Vogl}}, \bibinfo {author} {\bibfnamefont {B.}~\bibnamefont {Stiller}}, \bibinfo {author} {\bibfnamefont {G.}~\bibnamefont {Schunk}}, \bibinfo {author} {\bibfnamefont {D.~V.}\ \bibnamefont {Strekalov}}, \bibinfo {author} {\bibfnamefont {C.}~\bibnamefont {Marquardt}}, \bibinfo {author} {\bibfnamefont {J.~M.}\ \bibnamefont {Fink}}, \bibinfo {author} {\bibfnamefont {O.}~\bibnamefont {Painter}}, \emph {et~al.},\ }\enquote{\bibinfo {title} {Efficient microwave to optical photon conversion: an electro-optical realization},} \href@noop {} {\bibfield  {journal} {\bibinfo  {journal} {\emph {Optica}}}\ }\textbf {\bibinfo {volume} {3}},\ \bibinfo {pages} {597} (\bibinfo {year} {2016})\BibitemShut {NoStop}%
\bibitem [{\citenamefont {Pintus}\ \emph {et~al.}(2022{\natexlab{b}})\citenamefont {Pintus}, \citenamefont {Ranzani}, \citenamefont {Pinna}, \citenamefont {Huang}, \citenamefont {Gustafsson}, \citenamefont {Karinou}, \citenamefont {Casula}, \citenamefont {Shoji}, \citenamefont {Takamura}, \citenamefont {Mizumoto} \emph {et~al.}}]{pintus2022integrated}%
  \BibitemOpen
  \bibfield  {author} {\bibinfo {author} {\bibfnamefont {P.}~\bibnamefont {Pintus}}, \bibinfo {author} {\bibfnamefont {L.}~\bibnamefont {Ranzani}}, \bibinfo {author} {\bibfnamefont {S.}~\bibnamefont {Pinna}}, \bibinfo {author} {\bibfnamefont {D.}~\bibnamefont {Huang}}, \bibinfo {author} {\bibfnamefont {M.~V.}\ \bibnamefont {Gustafsson}}, \bibinfo {author} {\bibfnamefont {F.}~\bibnamefont {Karinou}}, \bibinfo {author} {\bibfnamefont {G.~A.}\ \bibnamefont {Casula}}, \bibinfo {author} {\bibfnamefont {Y.}~\bibnamefont {Shoji}}, \bibinfo {author} {\bibfnamefont {Y.}~\bibnamefont {Takamura}}, \bibinfo {author} {\bibfnamefont {T.}~\bibnamefont {Mizumoto}}, \emph {et~al.},\ }\enquote{\bibinfo {title} {An integrated magneto-optic modulator for cryogenic applications},} \href@noop {} {\bibfield  {journal} {\bibinfo  {journal} {\emph {Nat. Electron.}}}\ }\textbf {\bibinfo {volume} {5}},\ \bibinfo {pages} {604} (\bibinfo {year} {2022}{\natexlab{b}})\BibitemShut {NoStop}%
\bibitem [{\citenamefont {Zhu}\ \emph {et~al.}(2020)\citenamefont {Zhu}, \citenamefont {Zhang}, \citenamefont {Han}, \citenamefont {Zou}, \citenamefont {Zhong}, \citenamefont {Wang}, \citenamefont {Jiang},\ and\ \citenamefont {Tang}}]{zhu2020waveguide}%
  \BibitemOpen
  \bibfield  {author} {\bibinfo {author} {\bibfnamefont {N.}~\bibnamefont {Zhu}}, \bibinfo {author} {\bibfnamefont {X.}~\bibnamefont {Zhang}}, \bibinfo {author} {\bibfnamefont {X.}~\bibnamefont {Han}}, \bibinfo {author} {\bibfnamefont {C.-L.}\ \bibnamefont {Zou}}, \bibinfo {author} {\bibfnamefont {C.}~\bibnamefont {Zhong}}, \bibinfo {author} {\bibfnamefont {C.-H.}\ \bibnamefont {Wang}}, \bibinfo {author} {\bibfnamefont {L.}~\bibnamefont {Jiang}},\ and\ \bibinfo {author} {\bibfnamefont {H.~X.}\ \bibnamefont {Tang}},\ }\enquote{\bibinfo {title} {Waveguide cavity optomagnonics for microwave-to-optics conversion},} \href@noop {} {\bibfield  {journal} {\bibinfo  {journal} {\emph {Optica}}}\ }\textbf {\bibinfo {volume} {7}},\ \bibinfo {pages} {1291} (\bibinfo {year} {2020})\BibitemShut {NoStop}%
\bibitem [{\citenamefont {Shen}\ \emph {et~al.}(2022)\citenamefont {Shen}, \citenamefont {Xu}, \citenamefont {Zhang}, \citenamefont {Zhang}, \citenamefont {Wang}, \citenamefont {Chai}, \citenamefont {Zou}, \citenamefont {Guo},\ and\ \citenamefont {Dong}}]{shen2022coherent}%
  \BibitemOpen
  \bibfield  {author} {\bibinfo {author} {\bibfnamefont {Z.}~\bibnamefont {Shen}}, \bibinfo {author} {\bibfnamefont {G.-T.}\ \bibnamefont {Xu}}, \bibinfo {author} {\bibfnamefont {M.}~\bibnamefont {Zhang}}, \bibinfo {author} {\bibfnamefont {Y.-L.}\ \bibnamefont {Zhang}}, \bibinfo {author} {\bibfnamefont {Y.}~\bibnamefont {Wang}}, \bibinfo {author} {\bibfnamefont {C.-Z.}\ \bibnamefont {Chai}}, \bibinfo {author} {\bibfnamefont {C.-L.}\ \bibnamefont {Zou}}, \bibinfo {author} {\bibfnamefont {G.-C.}\ \bibnamefont {Guo}},\ and\ \bibinfo {author} {\bibfnamefont {C.-H.}\ \bibnamefont {Dong}},\ }\enquote{\bibinfo {title} {Coherent coupling between phonons, magnons, and photons},} \href@noop {} {\bibfield  {journal} {\bibinfo  {journal} {\emph {Phys. Rev. Lett.}}}\ }\textbf {\bibinfo {volume} {129}},\ \bibinfo {pages} {243601} (\bibinfo {year} {2022})\BibitemShut {NoStop}%
\bibitem [{\citenamefont {Chai}\ \emph {et~al.}(2022)\citenamefont {Chai}, \citenamefont {Shen}, \citenamefont {Zhang}, \citenamefont {Zhao}, \citenamefont {Guo}, \citenamefont {Zou},\ and\ \citenamefont {Dong}}]{chai2022single}%
  \BibitemOpen
  \bibfield  {author} {\bibinfo {author} {\bibfnamefont {C.-Z.}\ \bibnamefont {Chai}}, \bibinfo {author} {\bibfnamefont {Z.}~\bibnamefont {Shen}}, \bibinfo {author} {\bibfnamefont {Y.-L.}\ \bibnamefont {Zhang}}, \bibinfo {author} {\bibfnamefont {H.-Q.}\ \bibnamefont {Zhao}}, \bibinfo {author} {\bibfnamefont {G.-C.}\ \bibnamefont {Guo}}, \bibinfo {author} {\bibfnamefont {C.-L.}\ \bibnamefont {Zou}},\ and\ \bibinfo {author} {\bibfnamefont {C.-H.}\ \bibnamefont {Dong}},\ }\enquote{\bibinfo {title} {Single-sideband microwave-to-optical conversion in high-{Q} ferrimagnetic microspheres},} \href@noop {} {\bibfield  {journal} {\bibinfo  {journal} {\emph {Photon. Res.}}}\ }\textbf {\bibinfo {volume} {10}},\ \bibinfo {pages} {820} (\bibinfo {year} {2022})\BibitemShut {NoStop}%
\bibitem [{\citenamefont {Mirhosseini}\ \emph {et~al.}(2020)\citenamefont {Mirhosseini}, \citenamefont {Sipahigil}, \citenamefont {Kalaee},\ and\ \citenamefont {Painter}}]{mirhosseini2020superconducting}%
  \BibitemOpen
  \bibfield  {author} {\bibinfo {author} {\bibfnamefont {M.}~\bibnamefont {Mirhosseini}}, \bibinfo {author} {\bibfnamefont {A.}~\bibnamefont {Sipahigil}}, \bibinfo {author} {\bibfnamefont {M.}~\bibnamefont {Kalaee}},\ and\ \bibinfo {author} {\bibfnamefont {O.}~\bibnamefont {Painter}},\ }\enquote{\bibinfo {title} {Superconducting qubit to optical photon transduction},} \href@noop {} {\bibfield  {journal} {\bibinfo  {journal} {\emph {Nature}}}\ }\textbf {\bibinfo {volume} {588}},\ \bibinfo {pages} {599} (\bibinfo {year} {2020})\BibitemShut {NoStop}%
\bibitem [{\citenamefont {Zhao}\ \emph {et~al.}(2025)\citenamefont {Zhao}, \citenamefont {Chen}, \citenamefont {Kejriwal},\ and\ \citenamefont {Mirhosseini}}]{zhao2025quantum}%
  \BibitemOpen
  \bibfield  {author} {\bibinfo {author} {\bibfnamefont {H.}~\bibnamefont {Zhao}}, \bibinfo {author} {\bibfnamefont {W.~D.}\ \bibnamefont {Chen}}, \bibinfo {author} {\bibfnamefont {A.}~\bibnamefont {Kejriwal}},\ and\ \bibinfo {author} {\bibfnamefont {M.}~\bibnamefont {Mirhosseini}},\ }\enquote{\bibinfo {title} {Quantum-enabled microwave-to-optical transduction via silicon nanomechanics},} \href@noop {} {\bibfield  {journal} {\bibinfo  {journal} {\emph {Nat. Nanotechnol.}}}\ }\textbf {\bibinfo {volume} {20}},\ \bibinfo {pages} {602} (\bibinfo {year} {2025})\BibitemShut {NoStop}%
\bibitem [{\citenamefont {van Thiel}\ \emph {et~al.}(2025)\citenamefont {van Thiel}, \citenamefont {Weaver}, \citenamefont {Berto}, \citenamefont {Duivestein}, \citenamefont {Lemang}, \citenamefont {Schuurman}, \citenamefont {{\v{Z}}emli{\v{c}}ka}, \citenamefont {Hijazi}, \citenamefont {Bernasconi}, \citenamefont {Ferrer} \emph {et~al.}}]{van2025optical}%
  \BibitemOpen
  \bibfield  {author} {\bibinfo {author} {\bibfnamefont {T.}~\bibnamefont {van Thiel}}, \bibinfo {author} {\bibfnamefont {M.}~\bibnamefont {Weaver}}, \bibinfo {author} {\bibfnamefont {F.}~\bibnamefont {Berto}}, \bibinfo {author} {\bibfnamefont {P.}~\bibnamefont {Duivestein}}, \bibinfo {author} {\bibfnamefont {M.}~\bibnamefont {Lemang}}, \bibinfo {author} {\bibfnamefont {K.}~\bibnamefont {Schuurman}}, \bibinfo {author} {\bibfnamefont {M.}~\bibnamefont {{\v{Z}}emli{\v{c}}ka}}, \bibinfo {author} {\bibfnamefont {F.}~\bibnamefont {Hijazi}}, \bibinfo {author} {\bibfnamefont {A.}~\bibnamefont {Bernasconi}}, \bibinfo {author} {\bibfnamefont {C.}~\bibnamefont {Ferrer}}, \emph {et~al.},\ }\enquote{\bibinfo {title} {Optical readout of a superconducting qubit using a piezo-optomechanical transducer},} \href@noop {} {\bibfield  {journal} {\bibinfo  {journal} {\emph {Nat. Phys.}}}\ }\textbf {\bibinfo {volume} {21}},\ \bibinfo {pages} {401} (\bibinfo {year} {2025})\BibitemShut {NoStop}%
\bibitem [{\citenamefont {Andrews}\ \emph {et~al.}(2014)\citenamefont {Andrews}, \citenamefont {Peterson}, \citenamefont {Purdy}, \citenamefont {Cicak}, \citenamefont {Simmonds}, \citenamefont {Regal},\ and\ \citenamefont {Lehnert}}]{andrews2014bidirectional}%
  \BibitemOpen
  \bibfield  {author} {\bibinfo {author} {\bibfnamefont {R.~W.}\ \bibnamefont {Andrews}}, \bibinfo {author} {\bibfnamefont {R.~W.}\ \bibnamefont {Peterson}}, \bibinfo {author} {\bibfnamefont {T.~P.}\ \bibnamefont {Purdy}}, \bibinfo {author} {\bibfnamefont {K.}~\bibnamefont {Cicak}}, \bibinfo {author} {\bibfnamefont {R.~W.}\ \bibnamefont {Simmonds}}, \bibinfo {author} {\bibfnamefont {C.~A.}\ \bibnamefont {Regal}},\ and\ \bibinfo {author} {\bibfnamefont {K.~W.}\ \bibnamefont {Lehnert}},\ }\enquote{\bibinfo {title} {Bidirectional and efficient conversion between microwave and optical light},} \href@noop {} {\bibfield  {journal} {\bibinfo  {journal} {\emph {Nat. Phys.}}}\ }\textbf {\bibinfo {volume} {10}},\ \bibinfo {pages} {321} (\bibinfo {year} {2014})\BibitemShut {NoStop}%
\bibitem [{\citenamefont {Jiang}\ \emph {et~al.}(2020)\citenamefont {Jiang}, \citenamefont {Sarabalis}, \citenamefont {Dahmani}, \citenamefont {Patel}, \citenamefont {Mayor}, \citenamefont {McKenna}, \citenamefont {Van~Laer},\ and\ \citenamefont {Safavi-Naeini}}]{jiang2020efficient}%
  \BibitemOpen
  \bibfield  {author} {\bibinfo {author} {\bibfnamefont {W.}~\bibnamefont {Jiang}}, \bibinfo {author} {\bibfnamefont {C.~J.}\ \bibnamefont {Sarabalis}}, \bibinfo {author} {\bibfnamefont {Y.~D.}\ \bibnamefont {Dahmani}}, \bibinfo {author} {\bibfnamefont {R.~N.}\ \bibnamefont {Patel}}, \bibinfo {author} {\bibfnamefont {F.~M.}\ \bibnamefont {Mayor}}, \bibinfo {author} {\bibfnamefont {T.~P.}\ \bibnamefont {McKenna}}, \bibinfo {author} {\bibfnamefont {R.}~\bibnamefont {Van~Laer}},\ and\ \bibinfo {author} {\bibfnamefont {A.~H.}\ \bibnamefont {Safavi-Naeini}},\ }\enquote{\bibinfo {title} {Efficient bidirectional piezo-optomechanical transduction between microwave and optical frequency},} \href@noop {} {\bibfield  {journal} {\bibinfo  {journal} {\emph {Nat. Commun.}}}\ }\textbf {\bibinfo {volume} {11}},\ \bibinfo {pages} {1166} (\bibinfo {year} {2020})\BibitemShut {NoStop}%
\bibitem [{\citenamefont {Han}\ \emph {et~al.}(2020)\citenamefont {Han}, \citenamefont {Fu}, \citenamefont {Zhong}, \citenamefont {Zou}, \citenamefont {Xu}, \citenamefont {Sayem}, \citenamefont {Xu}, \citenamefont {Wang}, \citenamefont {Cheng}, \citenamefont {Jiang} \emph {et~al.}}]{han2020cavity}%
  \BibitemOpen
  \bibfield  {author} {\bibinfo {author} {\bibfnamefont {X.}~\bibnamefont {Han}}, \bibinfo {author} {\bibfnamefont {W.}~\bibnamefont {Fu}}, \bibinfo {author} {\bibfnamefont {C.}~\bibnamefont {Zhong}}, \bibinfo {author} {\bibfnamefont {C.-L.}\ \bibnamefont {Zou}}, \bibinfo {author} {\bibfnamefont {Y.}~\bibnamefont {Xu}}, \bibinfo {author} {\bibfnamefont {A.~A.}\ \bibnamefont {Sayem}}, \bibinfo {author} {\bibfnamefont {M.}~\bibnamefont {Xu}}, \bibinfo {author} {\bibfnamefont {S.}~\bibnamefont {Wang}}, \bibinfo {author} {\bibfnamefont {R.}~\bibnamefont {Cheng}}, \bibinfo {author} {\bibfnamefont {L.}~\bibnamefont {Jiang}}, \emph {et~al.},\ }\enquote{\bibinfo {title} {Cavity piezo-mechanics for superconducting-nanophotonic quantum interface},} \href@noop {} {\bibfield  {journal} {\bibinfo  {journal} {\emph {Nat. Commun.}}}\ }\textbf {\bibinfo {volume} {11}},\ \bibinfo {pages} {3237} (\bibinfo {year} {2020})\BibitemShut {NoStop}%
\bibitem [{\citenamefont {Weaver}\ \emph {et~al.}(2024)\citenamefont {Weaver}, \citenamefont {Duivestein}, \citenamefont {Bernasconi}, \citenamefont {Scharmer}, \citenamefont {Lemang}, \citenamefont {Thiel}, \citenamefont {Hijazi}, \citenamefont {Hensen}, \citenamefont {Gr{\"o}blacher},\ and\ \citenamefont {Stockill}}]{weaver2024integrated}%
  \BibitemOpen
  \bibfield  {author} {\bibinfo {author} {\bibfnamefont {M.~J.}\ \bibnamefont {Weaver}}, \bibinfo {author} {\bibfnamefont {P.}~\bibnamefont {Duivestein}}, \bibinfo {author} {\bibfnamefont {A.~C.}\ \bibnamefont {Bernasconi}}, \bibinfo {author} {\bibfnamefont {S.}~\bibnamefont {Scharmer}}, \bibinfo {author} {\bibfnamefont {M.}~\bibnamefont {Lemang}}, \bibinfo {author} {\bibfnamefont {T.~C.~v.}\ \bibnamefont {Thiel}}, \bibinfo {author} {\bibfnamefont {F.}~\bibnamefont {Hijazi}}, \bibinfo {author} {\bibfnamefont {B.}~\bibnamefont {Hensen}}, \bibinfo {author} {\bibfnamefont {S.}~\bibnamefont {Gr{\"o}blacher}},\ and\ \bibinfo {author} {\bibfnamefont {R.}~\bibnamefont {Stockill}},\ }\enquote{\bibinfo {title} {An integrated microwave-to-optics interface for scalable quantum computing},} \href@noop {} {\bibfield  {journal} {\bibinfo  {journal} {\emph {Nat. Nanotechnol.}}}\ }\textbf {\bibinfo {volume} {19}},\ \bibinfo {pages} {166} (\bibinfo {year} {2024})\BibitemShut {NoStop}%
\bibitem [{\citenamefont {Zhou}\ \emph {et~al.}(2024)\citenamefont {Zhou}, \citenamefont {Ruesink}, \citenamefont {Pavlovich}, \citenamefont {Behunin}, \citenamefont {Cheng}, \citenamefont {Gertler}, \citenamefont {Starbuck}, \citenamefont {Leenheer}, \citenamefont {Pomerene}, \citenamefont {Trotter} \emph {et~al.}}]{zhou2024electrically}%
  \BibitemOpen
  \bibfield  {author} {\bibinfo {author} {\bibfnamefont {Y.}~\bibnamefont {Zhou}}, \bibinfo {author} {\bibfnamefont {F.}~\bibnamefont {Ruesink}}, \bibinfo {author} {\bibfnamefont {M.}~\bibnamefont {Pavlovich}}, \bibinfo {author} {\bibfnamefont {R.}~\bibnamefont {Behunin}}, \bibinfo {author} {\bibfnamefont {H.}~\bibnamefont {Cheng}}, \bibinfo {author} {\bibfnamefont {S.}~\bibnamefont {Gertler}}, \bibinfo {author} {\bibfnamefont {A.~L.}\ \bibnamefont {Starbuck}}, \bibinfo {author} {\bibfnamefont {A.~J.}\ \bibnamefont {Leenheer}}, \bibinfo {author} {\bibfnamefont {A.~T.}\ \bibnamefont {Pomerene}}, \bibinfo {author} {\bibfnamefont {D.~C.}\ \bibnamefont {Trotter}}, \emph {et~al.},\ }\enquote{\bibinfo {title} {Electrically interfaced {Brillouin}-active waveguide for microwave photonic measurements},} \href@noop {} {\bibfield  {journal} {\bibinfo  {journal} {\emph {Nat. Commun.}}}\ }\textbf {\bibinfo {volume} {15}},\ \bibinfo {pages} {6796} (\bibinfo {year} {2024})\BibitemShut {NoStop}%
\bibitem [{\citenamefont {Arnold}\ \emph {et~al.}(2020)\citenamefont {Arnold}, \citenamefont {Wulf}, \citenamefont {Barzanjeh}, \citenamefont {Redchenko}, \citenamefont {Rueda}, \citenamefont {Hease}, \citenamefont {Hassani},\ and\ \citenamefont {Fink}}]{arnold2020converting}%
  \BibitemOpen
  \bibfield  {author} {\bibinfo {author} {\bibfnamefont {G.}~\bibnamefont {Arnold}}, \bibinfo {author} {\bibfnamefont {M.}~\bibnamefont {Wulf}}, \bibinfo {author} {\bibfnamefont {S.}~\bibnamefont {Barzanjeh}}, \bibinfo {author} {\bibfnamefont {E.}~\bibnamefont {Redchenko}}, \bibinfo {author} {\bibfnamefont {A.}~\bibnamefont {Rueda}}, \bibinfo {author} {\bibfnamefont {W.~J.}\ \bibnamefont {Hease}}, \bibinfo {author} {\bibfnamefont {F.}~\bibnamefont {Hassani}},\ and\ \bibinfo {author} {\bibfnamefont {J.~M.}\ \bibnamefont {Fink}},\ }\enquote{\bibinfo {title} {Converting microwave and telecom photons with a silicon photonic nanomechanical interface},} \href@noop {} {\bibfield  {journal} {\bibinfo  {journal} {\emph {Nat. Commun.}}}\ }\textbf {\bibinfo {volume} {11}},\ \bibinfo {pages} {4460} (\bibinfo {year} {2020})\BibitemShut {NoStop}%
\bibitem [{\citenamefont {Meesala}\ \emph {et~al.}(2024{\natexlab{a}})\citenamefont {Meesala}, \citenamefont {Wood}, \citenamefont {Lake}, \citenamefont {Chiappina}, \citenamefont {Zhong}, \citenamefont {Beyer}, \citenamefont {Shaw}, \citenamefont {Jiang},\ and\ \citenamefont {Painter}}]{meesala2024non}%
  \BibitemOpen
  \bibfield  {author} {\bibinfo {author} {\bibfnamefont {S.}~\bibnamefont {Meesala}}, \bibinfo {author} {\bibfnamefont {S.}~\bibnamefont {Wood}}, \bibinfo {author} {\bibfnamefont {D.}~\bibnamefont {Lake}}, \bibinfo {author} {\bibfnamefont {P.}~\bibnamefont {Chiappina}}, \bibinfo {author} {\bibfnamefont {C.}~\bibnamefont {Zhong}}, \bibinfo {author} {\bibfnamefont {A.~D.}\ \bibnamefont {Beyer}}, \bibinfo {author} {\bibfnamefont {M.~D.}\ \bibnamefont {Shaw}}, \bibinfo {author} {\bibfnamefont {L.}~\bibnamefont {Jiang}},\ and\ \bibinfo {author} {\bibfnamefont {O.}~\bibnamefont {Painter}},\ }\enquote{\bibinfo {title} {Non-classical microwave--optical photon pair generation with a chip-scale transducer},} \href@noop {} {\bibfield  {journal} {\bibinfo  {journal} {\emph {Nat. Phys.}}}\ }\textbf {\bibinfo {volume} {20}},\ \bibinfo {pages} {871} (\bibinfo {year} {2024}{\natexlab{a}})\BibitemShut {NoStop}%
\bibitem [{\citenamefont {Meesala}\ \emph {et~al.}(2024{\natexlab{b}})\citenamefont {Meesala}, \citenamefont {Lake}, \citenamefont {Wood}, \citenamefont {Chiappina}, \citenamefont {Zhong}, \citenamefont {Beyer}, \citenamefont {Shaw}, \citenamefont {Jiang},\ and\ \citenamefont {Painter}}]{meesala2024quantum}%
  \BibitemOpen
  \bibfield  {author} {\bibinfo {author} {\bibfnamefont {S.}~\bibnamefont {Meesala}}, \bibinfo {author} {\bibfnamefont {D.}~\bibnamefont {Lake}}, \bibinfo {author} {\bibfnamefont {S.}~\bibnamefont {Wood}}, \bibinfo {author} {\bibfnamefont {P.}~\bibnamefont {Chiappina}}, \bibinfo {author} {\bibfnamefont {C.}~\bibnamefont {Zhong}}, \bibinfo {author} {\bibfnamefont {A.~D.}\ \bibnamefont {Beyer}}, \bibinfo {author} {\bibfnamefont {M.~D.}\ \bibnamefont {Shaw}}, \bibinfo {author} {\bibfnamefont {L.}~\bibnamefont {Jiang}},\ and\ \bibinfo {author} {\bibfnamefont {O.}~\bibnamefont {Painter}},\ }\enquote{\bibinfo {title} {Quantum entanglement between optical and microwave photonic qubits},} \href@noop {} {\bibfield  {journal} {\bibinfo  {journal} {\emph {Phys. Rev. X}}}\ }\textbf {\bibinfo {volume} {14}},\ \bibinfo {pages} {031055} (\bibinfo {year} {2024}{\natexlab{b}})\BibitemShut {NoStop}%
\bibitem [{\citenamefont {Xie}\ \emph {et~al.}(2025)\citenamefont {Xie}, \citenamefont {Fukumori}, \citenamefont {Li},\ and\ \citenamefont {Faraon}}]{xie2025scalable}%
  \BibitemOpen
  \bibfield  {author} {\bibinfo {author} {\bibfnamefont {T.}~\bibnamefont {Xie}}, \bibinfo {author} {\bibfnamefont {R.}~\bibnamefont {Fukumori}}, \bibinfo {author} {\bibfnamefont {J.}~\bibnamefont {Li}},\ and\ \bibinfo {author} {\bibfnamefont {A.}~\bibnamefont {Faraon}},\ }\enquote{\bibinfo {title} {Scalable microwave-to-optical transducers at the single-photon level with spins},} \href@noop {} {\bibfield  {journal} {\bibinfo  {journal} {\emph {Nat. Phys.}}}\ }\textbf {\bibinfo {volume} {21}},\ \bibinfo {pages} {931} (\bibinfo {year} {2025})\BibitemShut {NoStop}%
\bibitem [{\citenamefont {Rochman}\ \emph {et~al.}(2023)\citenamefont {Rochman}, \citenamefont {Xie}, \citenamefont {Bartholomew}, \citenamefont {Schwab},\ and\ \citenamefont {Faraon}}]{rochman2023microwave}%
  \BibitemOpen
  \bibfield  {author} {\bibinfo {author} {\bibfnamefont {J.}~\bibnamefont {Rochman}}, \bibinfo {author} {\bibfnamefont {T.}~\bibnamefont {Xie}}, \bibinfo {author} {\bibfnamefont {J.~G.}\ \bibnamefont {Bartholomew}}, \bibinfo {author} {\bibfnamefont {K.}~\bibnamefont {Schwab}},\ and\ \bibinfo {author} {\bibfnamefont {A.}~\bibnamefont {Faraon}},\ }\enquote{\bibinfo {title} {Microwave-to-optical transduction with erbium ions coupled to planar photonic and superconducting resonators},} \href@noop {} {\bibfield  {journal} {\bibinfo  {journal} {\emph {Nat. Commun.}}}\ }\textbf {\bibinfo {volume} {14}},\ \bibinfo {pages} {1153} (\bibinfo {year} {2023})\BibitemShut {NoStop}%
\bibitem [{\citenamefont {Nicolas}\ \emph {et~al.}(2023)\citenamefont {Nicolas}, \citenamefont {Businger}, \citenamefont {Sanchez~Mejia}, \citenamefont {Tiranov}, \citenamefont {Chaneli{\`e}re}, \citenamefont {Lafitte-Houssat}, \citenamefont {Ferrier}, \citenamefont {Goldner},\ and\ \citenamefont {Afzelius}}]{nicolas2023coherent}%
  \BibitemOpen
  \bibfield  {author} {\bibinfo {author} {\bibfnamefont {L.}~\bibnamefont {Nicolas}}, \bibinfo {author} {\bibfnamefont {M.}~\bibnamefont {Businger}}, \bibinfo {author} {\bibfnamefont {T.}~\bibnamefont {Sanchez~Mejia}}, \bibinfo {author} {\bibfnamefont {A.}~\bibnamefont {Tiranov}}, \bibinfo {author} {\bibfnamefont {T.}~\bibnamefont {Chaneli{\`e}re}}, \bibinfo {author} {\bibfnamefont {E.}~\bibnamefont {Lafitte-Houssat}}, \bibinfo {author} {\bibfnamefont {A.}~\bibnamefont {Ferrier}}, \bibinfo {author} {\bibfnamefont {P.}~\bibnamefont {Goldner}},\ and\ \bibinfo {author} {\bibfnamefont {M.}~\bibnamefont {Afzelius}},\ }\enquote{\bibinfo {title} {Coherent optical-microwave interface for manipulation of low-field electronic clock transitions in ${}^{171}${Yb}${}^{3+}$:{Y}${}_{2}${SiO}${}_5$},} \href@noop {} {\bibfield  {journal} {\bibinfo  {journal} {\emph {npj Quantum Info.}}}\ }\textbf {\bibinfo {volume} {9}},\ \bibinfo {pages} {21} (\bibinfo {year} {2023})\BibitemShut {NoStop}%
\bibitem [{\citenamefont {Xie}\ \emph {et~al.}(2021)\citenamefont {Xie}, \citenamefont {Rochman}, \citenamefont {Bartholomew}, \citenamefont {Ruskuc}, \citenamefont {Kindem}, \citenamefont {Craiciu}, \citenamefont {Thiel}, \citenamefont {Cone},\ and\ \citenamefont {Faraon}}]{xie2021characterization}%
  \BibitemOpen
  \bibfield  {author} {\bibinfo {author} {\bibfnamefont {T.}~\bibnamefont {Xie}}, \bibinfo {author} {\bibfnamefont {J.}~\bibnamefont {Rochman}}, \bibinfo {author} {\bibfnamefont {J.~G.}\ \bibnamefont {Bartholomew}}, \bibinfo {author} {\bibfnamefont {A.}~\bibnamefont {Ruskuc}}, \bibinfo {author} {\bibfnamefont {J.~M.}\ \bibnamefont {Kindem}}, \bibinfo {author} {\bibfnamefont {I.}~\bibnamefont {Craiciu}}, \bibinfo {author} {\bibfnamefont {C.~W.}\ \bibnamefont {Thiel}}, \bibinfo {author} {\bibfnamefont {R.~L.}\ \bibnamefont {Cone}},\ and\ \bibinfo {author} {\bibfnamefont {A.}~\bibnamefont {Faraon}},\ }\enquote{\bibinfo {title} {Characterization of {Er${}^{3+}$:YVO${}_4$} for microwave to optical transduction},} \href@noop {} {\bibfield  {journal} {\bibinfo  {journal} {\emph {Phys. Rev. B}}}\ }\textbf {\bibinfo {volume} {104}},\ \bibinfo {pages} {054111} (\bibinfo {year} {2021})\BibitemShut {NoStop}%
\bibitem [{\citenamefont {Han}\ \emph {et~al.}(2021)\citenamefont {Han}, \citenamefont {Fu}, \citenamefont {Zou}, \citenamefont {Jiang},\ and\ \citenamefont {Tang}}]{han2021microwave}%
  \BibitemOpen
  \bibfield  {author} {\bibinfo {author} {\bibfnamefont {X.}~\bibnamefont {Han}}, \bibinfo {author} {\bibfnamefont {W.}~\bibnamefont {Fu}}, \bibinfo {author} {\bibfnamefont {C.-L.}\ \bibnamefont {Zou}}, \bibinfo {author} {\bibfnamefont {L.}~\bibnamefont {Jiang}},\ and\ \bibinfo {author} {\bibfnamefont {H.~X.}\ \bibnamefont {Tang}},\ }\enquote{\bibinfo {title} {Microwave-optical quantum frequency conversion},} \href@noop {} {\bibfield  {journal} {\bibinfo  {journal} {\emph {Optica}}}\ }\textbf {\bibinfo {volume} {8}},\ \bibinfo {pages} {1050} (\bibinfo {year} {2021})\BibitemShut {NoStop}%
\bibitem [{\citenamefont {Boyd}(2020)}]{boydNLO}%
  \BibitemOpen
  \bibfield  {author} {\bibinfo {author} {\bibfnamefont {R.~W.}\ \bibnamefont {Boyd}},\ }\href@noop {} {\emph {\bibinfo {title} {Nonlinear optics}}},\ \bibinfo {edition} {4th}\ ed.\ (\bibinfo  {publisher} {Academic Press},\ \bibinfo {year} {2020})\BibitemShut {NoStop}%
\bibitem [{\citenamefont {Zhou}\ \emph {et~al.}(2025)\citenamefont {Zhou}, \citenamefont {Shen}, \citenamefont {Li}, \citenamefont {Yang}, \citenamefont {Xie},\ and\ \citenamefont {Tang}}]{zhou2025high}%
  \BibitemOpen
  \bibfield  {author} {\bibinfo {author} {\bibfnamefont {Y.}~\bibnamefont {Zhou}}, \bibinfo {author} {\bibfnamefont {M.}~\bibnamefont {Shen}}, \bibinfo {author} {\bibfnamefont {C.}~\bibnamefont {Li}}, \bibinfo {author} {\bibfnamefont {L.}~\bibnamefont {Yang}}, \bibinfo {author} {\bibfnamefont {J.}~\bibnamefont {Xie}},\ and\ \bibinfo {author} {\bibfnamefont {H.~X.}\ \bibnamefont {Tang}},\ }\enquote{\bibinfo {title} {High-efficiency, cryogenic-compatible grating couplers on an {AlN}-on-sapphire platform through bottom-side coupling},} \href@noop {} {\bibfield  {journal} {\bibinfo  {journal} {\emph {Opt. Lett.}}}\ }\textbf {\bibinfo {volume} {50}},\ \bibinfo {pages} {742} (\bibinfo {year} {2025})\BibitemShut {NoStop}%
\bibitem [{\citenamefont {Liao}\ \emph {et~al.}(2020)\citenamefont {Liao}, \citenamefont {Hu}, \citenamefont {Gan}, \citenamefont {Liu}, \citenamefont {Wu}, \citenamefont {Fan}, \citenamefont {Feng}, \citenamefont {Lu}, \citenamefont {Liu},\ and\ \citenamefont {Gong}}]{liao2020photonic}%
  \BibitemOpen
  \bibfield  {author} {\bibinfo {author} {\bibfnamefont {K.}~\bibnamefont {Liao}}, \bibinfo {author} {\bibfnamefont {X.}~\bibnamefont {Hu}}, \bibinfo {author} {\bibfnamefont {T.}~\bibnamefont {Gan}}, \bibinfo {author} {\bibfnamefont {Q.}~\bibnamefont {Liu}}, \bibinfo {author} {\bibfnamefont {Z.}~\bibnamefont {Wu}}, \bibinfo {author} {\bibfnamefont {C.}~\bibnamefont {Fan}}, \bibinfo {author} {\bibfnamefont {X.}~\bibnamefont {Feng}}, \bibinfo {author} {\bibfnamefont {C.}~\bibnamefont {Lu}}, \bibinfo {author} {\bibfnamefont {Y.-c.}\ \bibnamefont {Liu}},\ and\ \bibinfo {author} {\bibfnamefont {Q.}~\bibnamefont {Gong}},\ }\enquote{\bibinfo {title} {Photonic molecule quantum optics},} \href@noop {} {\bibfield  {journal} {\bibinfo  {journal} {\emph {Adv. Opt. Photon.}}}\ }\textbf {\bibinfo {volume} {12}},\ \bibinfo {pages} {60} (\bibinfo {year} {2020})\BibitemShut {NoStop}%
\bibitem [{\citenamefont {Fu}\ \emph {et~al.}(2021)\citenamefont {Fu}, \citenamefont {Xu}, \citenamefont {Liu}, \citenamefont {Zou}, \citenamefont {Zhong}, \citenamefont {Han}, \citenamefont {Shen}, \citenamefont {Xu}, \citenamefont {Cheng}, \citenamefont {Wang} \emph {et~al.}}]{fu2021cavity}%
  \BibitemOpen
  \bibfield  {author} {\bibinfo {author} {\bibfnamefont {W.}~\bibnamefont {Fu}}, \bibinfo {author} {\bibfnamefont {M.}~\bibnamefont {Xu}}, \bibinfo {author} {\bibfnamefont {X.}~\bibnamefont {Liu}}, \bibinfo {author} {\bibfnamefont {C.-L.}\ \bibnamefont {Zou}}, \bibinfo {author} {\bibfnamefont {C.}~\bibnamefont {Zhong}}, \bibinfo {author} {\bibfnamefont {X.}~\bibnamefont {Han}}, \bibinfo {author} {\bibfnamefont {M.}~\bibnamefont {Shen}}, \bibinfo {author} {\bibfnamefont {Y.}~\bibnamefont {Xu}}, \bibinfo {author} {\bibfnamefont {R.}~\bibnamefont {Cheng}}, \bibinfo {author} {\bibfnamefont {S.}~\bibnamefont {Wang}}, \emph {et~al.},\ }\enquote{\bibinfo {title} {Cavity electro-optic circuit for microwave-to-optical conversion in the quantum ground state},} \href@noop {} {\bibfield  {journal} {\bibinfo  {journal} {\emph {Phys. Rev. A}}}\ }\textbf {\bibinfo {volume} {103}},\ \bibinfo {pages} {053504} (\bibinfo {year} {2021})\BibitemShut {NoStop}%
\bibitem [{\citenamefont {Liu}\ \emph {et~al.}(2023)\citenamefont {Liu}, \citenamefont {Bruch},\ and\ \citenamefont {Tang}}]{liu2023aluminum}%
  \BibitemOpen
  \bibfield  {author} {\bibinfo {author} {\bibfnamefont {X.}~\bibnamefont {Liu}}, \bibinfo {author} {\bibfnamefont {A.~W.}\ \bibnamefont {Bruch}},\ and\ \bibinfo {author} {\bibfnamefont {H.~X.}\ \bibnamefont {Tang}},\ }\enquote{\bibinfo {title} {Aluminum nitride photonic integrated circuits: from piezo-optomechanics to nonlinear optics},} \href@noop {} {\bibfield  {journal} {\bibinfo  {journal} {\emph {Adv. Opt. Photon.}}}\ }\textbf {\bibinfo {volume} {15}},\ \bibinfo {pages} {236} (\bibinfo {year} {2023})\BibitemShut {NoStop}%
\bibitem [{\citenamefont {Li}\ \emph {et~al.}(2025)\citenamefont {Li}, \citenamefont {Xu}, \citenamefont {Wu}, \citenamefont {Pace}, \citenamefont {LaHaye}, \citenamefont {Senatore},\ and\ \citenamefont {Tang}}]{li2025fast}%
  \BibitemOpen
  \bibfield  {author} {\bibinfo {author} {\bibfnamefont {C.}~\bibnamefont {Li}}, \bibinfo {author} {\bibfnamefont {Y.}~\bibnamefont {Xu}}, \bibinfo {author} {\bibfnamefont {Y.}~\bibnamefont {Wu}}, \bibinfo {author} {\bibfnamefont {M.~C.~C.}\ \bibnamefont {Pace}}, \bibinfo {author} {\bibfnamefont {M.~D.}\ \bibnamefont {LaHaye}}, \bibinfo {author} {\bibfnamefont {M.}~\bibnamefont {Senatore}},\ and\ \bibinfo {author} {\bibfnamefont {H.~X.}\ \bibnamefont {Tang}},\ }\enquote{\bibinfo {title} {Fast recovery of niobium-based superconducting resonators after laser illumination},} \href@noop {} {\bibfield  {journal} {\bibinfo  {journal} {\emph {arXiv:2507.16082}}}\ } (\bibinfo {year} {2025})\BibitemShut {NoStop}%
\bibitem [{\citenamefont {Liu}\ \emph {et~al.}(2015)\citenamefont {Liu}, \citenamefont {Sun}, \citenamefont {Xiong}, \citenamefont {Niu}, \citenamefont {Hao}, \citenamefont {Han},\ and\ \citenamefont {Luo}}]{liu2015smooth}%
  \BibitemOpen
  \bibfield  {author} {\bibinfo {author} {\bibfnamefont {X.}~\bibnamefont {Liu}}, \bibinfo {author} {\bibfnamefont {C.}~\bibnamefont {Sun}}, \bibinfo {author} {\bibfnamefont {B.}~\bibnamefont {Xiong}}, \bibinfo {author} {\bibfnamefont {L.}~\bibnamefont {Niu}}, \bibinfo {author} {\bibfnamefont {Z.}~\bibnamefont {Hao}}, \bibinfo {author} {\bibfnamefont {Y.}~\bibnamefont {Han}},\ and\ \bibinfo {author} {\bibfnamefont {Y.}~\bibnamefont {Luo}},\ }\enquote{\bibinfo {title} {Smooth etching of epitaxially grown {AlN} film by {Cl${}_2$/BCl${}_3$/Ar}-based inductively coupled plasma},} \href@noop {} {\bibfield  {journal} {\bibinfo  {journal} {\emph {Vacuum}}}\ }\textbf {\bibinfo {volume} {116}},\ \bibinfo {pages} {158} (\bibinfo {year} {2015})\BibitemShut {NoStop}%
\bibitem [{\citenamefont {Macklin}\ \emph {et~al.}(2015)\citenamefont {Macklin}, \citenamefont {O’brien}, \citenamefont {Hover}, \citenamefont {Schwartz}, \citenamefont {Bolkhovsky}, \citenamefont {Zhang}, \citenamefont {Oliver},\ and\ \citenamefont {Siddiqi}}]{macklin2015near}%
  \BibitemOpen
  \bibfield  {author} {\bibinfo {author} {\bibfnamefont {C.}~\bibnamefont {Macklin}}, \bibinfo {author} {\bibfnamefont {K.}~\bibnamefont {O’brien}}, \bibinfo {author} {\bibfnamefont {D.}~\bibnamefont {Hover}}, \bibinfo {author} {\bibfnamefont {M.}~\bibnamefont {Schwartz}}, \bibinfo {author} {\bibfnamefont {V.}~\bibnamefont {Bolkhovsky}}, \bibinfo {author} {\bibfnamefont {X.}~\bibnamefont {Zhang}}, \bibinfo {author} {\bibfnamefont {W.}~\bibnamefont {Oliver}},\ and\ \bibinfo {author} {\bibfnamefont {I.}~\bibnamefont {Siddiqi}},\ }\enquote{\bibinfo {title} {A near--quantum-limited {Josephson} traveling-wave parametric amplifier},} \href@noop {} {\bibfield  {journal} {\bibinfo  {journal} {\emph {Science}}}\ }\textbf {\bibinfo {volume} {350}},\ \bibinfo {pages} {307} (\bibinfo {year} {2015})\BibitemShut {NoStop}%
\bibitem [{\citenamefont {Xu}\ \emph {et~al.}(2020)\citenamefont {Xu}, \citenamefont {Han}, \citenamefont {Zou}, \citenamefont {Fu}, \citenamefont {Xu}, \citenamefont {Zhong}, \citenamefont {Jiang},\ and\ \citenamefont {Tang}}]{xu2020radiative}%
  \BibitemOpen
  \bibfield  {author} {\bibinfo {author} {\bibfnamefont {M.}~\bibnamefont {Xu}}, \bibinfo {author} {\bibfnamefont {X.}~\bibnamefont {Han}}, \bibinfo {author} {\bibfnamefont {C.-L.}\ \bibnamefont {Zou}}, \bibinfo {author} {\bibfnamefont {W.}~\bibnamefont {Fu}}, \bibinfo {author} {\bibfnamefont {Y.}~\bibnamefont {Xu}}, \bibinfo {author} {\bibfnamefont {C.}~\bibnamefont {Zhong}}, \bibinfo {author} {\bibfnamefont {L.}~\bibnamefont {Jiang}},\ and\ \bibinfo {author} {\bibfnamefont {H.~X.}\ \bibnamefont {Tang}},\ }\enquote{\bibinfo {title} {Radiative cooling of a superconducting resonator},} \href@noop {} {\bibfield  {journal} {\bibinfo  {journal} {\emph {Phys. Rev. Lett.}}}\ }\textbf {\bibinfo {volume} {124}},\ \bibinfo {pages} {033602} (\bibinfo {year} {2020})\BibitemShut {NoStop}%
\bibitem [{\citenamefont {Duan}\ \emph {et~al.}(2001)\citenamefont {Duan}, \citenamefont {Lukin}, \citenamefont {Cirac},\ and\ \citenamefont {Zoller}}]{duan2001long}%
  \BibitemOpen
  \bibfield  {author} {\bibinfo {author} {\bibfnamefont {L.-M.}\ \bibnamefont {Duan}}, \bibinfo {author} {\bibfnamefont {M.~D.}\ \bibnamefont {Lukin}}, \bibinfo {author} {\bibfnamefont {J.~I.}\ \bibnamefont {Cirac}},\ and\ \bibinfo {author} {\bibfnamefont {P.}~\bibnamefont {Zoller}},\ }\enquote{\bibinfo {title} {Long-distance quantum communication with atomic ensembles and linear optics},} \href@noop {} {\bibfield  {journal} {\bibinfo  {journal} {\emph {Nature}}}\ }\textbf {\bibinfo {volume} {414}},\ \bibinfo {pages} {413} (\bibinfo {year} {2001})\BibitemShut {NoStop}%
\bibitem [{\citenamefont {Zhong}\ \emph {et~al.}(2020)\citenamefont {Zhong}, \citenamefont {Wang}, \citenamefont {Zou}, \citenamefont {Zhang}, \citenamefont {Han}, \citenamefont {Fu}, \citenamefont {Xu}, \citenamefont {Shankar}, \citenamefont {Devoret}, \citenamefont {Tang} \emph {et~al.}}]{zhong2020proposal}%
  \BibitemOpen
  \bibfield  {author} {\bibinfo {author} {\bibfnamefont {C.}~\bibnamefont {Zhong}}, \bibinfo {author} {\bibfnamefont {Z.}~\bibnamefont {Wang}}, \bibinfo {author} {\bibfnamefont {C.}~\bibnamefont {Zou}}, \bibinfo {author} {\bibfnamefont {M.}~\bibnamefont {Zhang}}, \bibinfo {author} {\bibfnamefont {X.}~\bibnamefont {Han}}, \bibinfo {author} {\bibfnamefont {W.}~\bibnamefont {Fu}}, \bibinfo {author} {\bibfnamefont {M.}~\bibnamefont {Xu}}, \bibinfo {author} {\bibfnamefont {S.}~\bibnamefont {Shankar}}, \bibinfo {author} {\bibfnamefont {M.~H.}\ \bibnamefont {Devoret}}, \bibinfo {author} {\bibfnamefont {H.~X.}\ \bibnamefont {Tang}}, \emph {et~al.},\ }\enquote{\bibinfo {title} {Proposal for heralded generation and detection of entangled microwave--optical-photon pairs},} \href@noop {} {\bibfield  {journal} {\bibinfo  {journal} {\emph {Phys. Rev. Lett.}}}\ }\textbf {\bibinfo {volume} {124}},\ \bibinfo {pages} {010511} (\bibinfo {year} {2020})\BibitemShut {NoStop}%
\bibitem [{\citenamefont {Rueda}\ \emph {et~al.}(2019)\citenamefont {Rueda}, \citenamefont {Hease}, \citenamefont {Barzanjeh},\ and\ \citenamefont {Fink}}]{rueda2019electro}%
  \BibitemOpen
  \bibfield  {author} {\bibinfo {author} {\bibfnamefont {A.}~\bibnamefont {Rueda}}, \bibinfo {author} {\bibfnamefont {W.}~\bibnamefont {Hease}}, \bibinfo {author} {\bibfnamefont {S.}~\bibnamefont {Barzanjeh}},\ and\ \bibinfo {author} {\bibfnamefont {J.~M.}\ \bibnamefont {Fink}},\ }\enquote{\bibinfo {title} {Electro-optic entanglement source for microwave to telecom quantum state transfer},} \href@noop {} {\bibfield  {journal} {\bibinfo  {journal} {\emph {npj Quantum Info.}}}\ }\textbf {\bibinfo {volume} {5}},\ \bibinfo {pages} {108} (\bibinfo {year} {2019})\BibitemShut {NoStop}%
\bibitem [{\citenamefont {Wolf}\ \emph {et~al.}(2007)\citenamefont {Wolf}, \citenamefont {P{\'e}rez-Garc{\'\i}a},\ and\ \citenamefont {Giedke}}]{wolf2007quantum}%
  \BibitemOpen
  \bibfield  {author} {\bibinfo {author} {\bibfnamefont {M.~M.}\ \bibnamefont {Wolf}}, \bibinfo {author} {\bibfnamefont {D.}~\bibnamefont {P{\'e}rez-Garc{\'\i}a}},\ and\ \bibinfo {author} {\bibfnamefont {G.}~\bibnamefont {Giedke}},\ }\enquote{\bibinfo {title} {Quantum capacities of bosonic channels},} \href@noop {} {\bibfield  {journal} {\bibinfo  {journal} {\emph {Phys. Rev. Lett.}}}\ }\textbf {\bibinfo {volume} {98}},\ \bibinfo {pages} {130501} (\bibinfo {year} {2007})\BibitemShut {NoStop}%
\bibitem [{\citenamefont {Gerry}\ and\ \citenamefont {Knight}(2023)}]{gerry2023introductory}%
  \BibitemOpen
  \bibfield  {author} {\bibinfo {author} {\bibfnamefont {C.~C.}\ \bibnamefont {Gerry}}\ and\ \bibinfo {author} {\bibfnamefont {P.~L.}\ \bibnamefont {Knight}},\ }\href@noop {} {\emph {\bibinfo {title} {Introductory quantum optics}}}\ (\bibinfo  {publisher} {Cambridge University Press},\ \bibinfo {year} {2023})\BibitemShut {NoStop}%
\end{thebibliography}
\end{document}